\documentclass[aps,prb,twocolumn,superscriptaddress,showpacs,floatfix,longbibliography]{revtex4-2}

\usepackage{amsmath,amssymb,amsfonts}
\usepackage{graphicx}
\graphicspath{{Figs/}}
\usepackage{bm}
\usepackage{xcolor}
\usepackage{hyperref}
\hypersetup{colorlinks=true,linkcolor=blue,citecolor=blue,urlcolor=blue}

\newcommand{\eqn}[1]{Eq.~(\ref{#1})}
\newcommand{\fig}[1]{Fig.~\ref{#1}}
\newcommand{\sect}[1]{Sec.~\ref{#1}}

\newcommand{\up}{\uparrow}
\newcommand{\dn}{\downarrow}

\begin{document}

\title{Exact Gaussian pinning of a relative-phase mode and
interaction-driven detuning from the pinned frequency}

\author{Yogeshwar Prasad}
\email{yogeshwar@snu.ac.kr}
\affiliation{Center for Condensed Matter Theory,
  Department of Physics,
  Indian Institute of Science,
  Bangalore 560012, India}
\affiliation{Department of Physics,
  Hanyang University,
  Seoul 04763, Korea}
\affiliation{Research Institute of Basic Sciences,
  Seoul National University,
  Seoul 08826, Korea}

\date{\today}

\begin{abstract}
Hybridization between two identical fermionic layers or components produces a rigid bonding--antibonding splitting $2t_h$. We show that in a paired state this scale enters the collective dynamics exactly within Gaussian fluctuation theory: the copy-odd relative-phase kernel vanishes at $\omega=2t_h$, with the underlying parity-pair identity holding pointwise in momentum. Within a broad class of conventionally paired systems, the result is independent of chemical potential, band structure, orbital content, and dimensionality. It follows from copy-pseudospin Larmor precession generated by intercopy hopping together with the Goldstone--Ward condition imposed by superconducting self-consistency. Pairing also isolates $2t_h$ from quasiparticle excitations, with thermal scattering below and pair creation above it whenever the gap is finite. At particle-hole symmetry, where the copy-odd amplitude decouples, the kernel zero becomes an isolated relative-phase pole at $\Omega_L=2t_h$, and the layer-imbalance response is a single Gaussian mode exhausting its Gaussian first moment; the corresponding full-Hamiltonian first moment obeys an exact operator sum rule. This pinning is not protected beyond Gaussian order. An exact interaction torque destroys the Larmor closure, while an exactly solvable rung and exact diagonalization of the attractive Hubbard model on a periodic $N_s=12$, $z=3$ bilayer honeycomb cluster show strong detuning. At $|U|=4t$, the dominant layer-imbalance excitation occurs at $0.467(2t_h)$ while retaining $98.6\%$ of the zeroth-moment spectral weight but only $92.4\%$ of the exact first moment. Thus the Gaussian result provides an exact reference frequency whose many-body renormalization can be substantial while a dominant layer-odd excitation survives.
\end{abstract}

\maketitle

\setcounter{topnumber}{1}
\setcounter{bottomnumber}{0}
\setcounter{totalnumber}{1}
\renewcommand{\topfraction}{0.9}
\renewcommand{\textfraction}{0.08}
\renewcommand{\floatpagefraction}{0.9}
\section{Introduction}
\label{sec:intro}

Relative-phase oscillations are a characteristic collective degree of
freedom of multicomponent superconductors. Since Leggett's original
analysis~\cite{Leggett1966}, they have been developed through
effective-action approaches~\cite{Sharapov2002}, strong-coupling
multiband models~\cite{Burnell2010}, and optical-response
theory~\cite{Kamatani2022}. Experimental signatures include the Raman
observation of the Leggett mode in
MgB$_2$~\cite{Blumberg2007,CeaBenfatto2016} and transport signatures in
a proximitized Dirac semimetal~\cite{Cuozzo2024}. Superconducting
bilayers add a distinct ingredient: single-particle interlayer hopping
directly hybridizes the two components and introduces a microscopic
bonding--antibonding scale into the relative-phase problem. Bilayer
phase modes~\cite{Hackner2023} and their accessibility to linear
spectroscopy in two dimensions~\cite{Levitan2024} have therefore
attracted renewed attention, while interacting bilayers with tunable
interlayer hopping are being explored as a route to controlling $T_c$
and the superfluid stiffness~\cite{Zhang2025bilayer,Fontenele2024}.
Attraction can also drive a superfluid from a bilayer band
insulator~\cite{PrasadShenoyPRA89}, whose finite-temperature
correlations have been studied by determinant quantum Monte
Carlo~\cite{Prasad2022}. These developments raise a sharper question:
can the relative-phase scale be fixed directly by the microscopic
interlayer hybridization, and if so, how robust is that pinning once
interactions are treated beyond Gaussian theory? We address this
question in a neutral short-range-attraction model. In a charged
electronic bilayer, long-range Coulomb interactions additionally dress
the layer-antisymmetric phase sector~\cite{Hackner2023}, an effect
outside the present Hamiltonian.

Closely related physics was identified previously in double-layer
superconductors. Forsthofer, Kind and Keller~\cite{Forsthofer1996}
studied the electronic polarization of a hybridized double layer within
a conserving RPA and, in the weak-coupling neutral limit with
momentum-independent hopping, obtained
$\omega_P^2=(2t)^2+\omega_0^2$, where $\omega_0$ is the Leggett
contribution. In the absence of interlayer pair transfer their result
reduces to a resonance at $2t$, below the pair-breaking threshold,
interpreted as collective pair tunneling between the layers. Our model
lies in this $\omega_0\to0$ sector. The advance pursued here is not the
existence of the $2t_h$ resonance, but the pointwise Gaussian identity
that fixes it, the structural conditions under which that identity is
general, and the manner in which the full interaction breaks the
pinning. In the particle-hole-symmetric form of \eqn{eqn:H} the Hartree
shift entering Ref.~\cite{Forsthofer1996} vanishes, so the
corresponding Gaussian scale is the bare $2t_h$.

\begin{figure}[t]
  \centerline{\includegraphics[width=\linewidth]{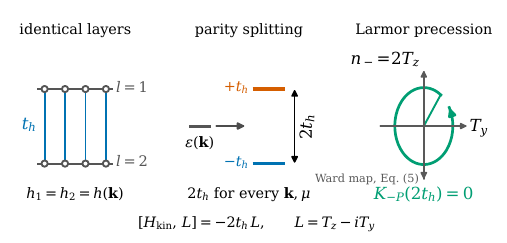}}
  \caption{Mechanism. Two identical layers $h_1=h_2=h(\mathbf k)$
  coupled by a momentum-independent $t_h$ (left) form bonding and
  antibonding parity partners split by the same $2t_h$ for every
  $\mathbf k$ and chemical potential (centre). In copy space the
  interlayer hopping acts as a uniform pseudomagnetic field, so the
  transverse copy pseudospin $L=T_z-iT_y$ precesses at the Larmor
  frequency $2t_h$ [\eqn{eqn:larmor}]. Superconducting
  self-consistency independently fixes the common-phase channel at zero
  frequency through the Goldstone--Ward condition. The rigid parity
  splitting then transfers this zero-frequency anchor to the copy-odd
  phase channel, giving the exact Gaussian kernel zero
  $K_{-P}(2t_h)=0$ through the pointwise identity \eqn{eqn:pointwise}
  (right). At particle-hole symmetry, where the copy-odd amplitude
  decouples, this kernel zero becomes the isolated relative-phase pole
  $\Omega_L=2t_h$.}
  \label{fig:mechanism}
\end{figure}

Here, however, $2t_h$ is not merely a weak-coupling solution of an RPA
equation. Momentum-independent intercopy hopping acts as a uniform
pseudomagnetic field in copy space, so that every bonding--antibonding
pair is split by the same $2t_h$ throughout the Brillouin zone.
Superconducting self-consistency independently fixes the common-phase
fluctuation at zero frequency through the U(1) Goldstone--Ward
condition. Because the copy splitting is rigid, this zero-frequency
phase structure is reproduced in the copy-odd channel at the finite
frequency $2t_h$. The underlying parity-pair identity holds pointwise in momentum, and
after summation the gap equation converts it into an exact Gaussian
kernel zero at $\omega=2t_h$, independent of interaction strength and
chemical potential. This mechanism does not rely on the
honeycomb dispersion: it applies to any pair of identical hybridized
copies, with arbitrary band structure, orbital content and
dimensionality, within the conventional bandwise singlet-pairing class
specified below (\sect{sec:general}). Its
isolation from the continuum follows from a pair of universal
inequalities that hold on the self-consistent paired saddle. At
particle-hole symmetry the layer-imbalance response is a single pole
exhausting that channel's f-sum rule. Our construction is the conserving Gaussian (RPA) response
of a paired state in the sense of
Refs.~\cite{Anderson1958,BaymKadanoff1961,Baym1962}. We take the
simplest bilayer realization, the attractive Hubbard model on an
$AA$-stacked bilayer honeycomb lattice;
Figs.~\ref{fig:mechanism}--\ref{fig:ed} summarize the mechanism,
the exactness and the physical response.

\section{Model and mean-field saddle}
\label{sec:model}

We consider the attractive Hubbard Hamiltonian
\begin{equation}
\begin{split}
  H = &-t\!\!\sum_{\langle ij\rangle,l,\sigma}\!\!
        \left(c^\dagger_{il\sigma}c_{jl\sigma}+{\rm H.c.}\right)
      - t_h\sum_{i,\sigma}\left(c^\dagger_{i1\sigma}c_{i2\sigma}+{\rm H.c.}\right)\\
      &- |U|\sum_{il}\Big(n_{il\up}-\tfrac12\Big)\Big(n_{il\dn}-\tfrac12\Big)
      - \mu\sum_{il\sigma} n_{il\sigma},
\end{split}
  \label{eqn:H}
\end{equation}
on the $AA$-stacked bilayer honeycomb lattice, where $l=1,2$ labels the
layers, $t$ is the intralayer nearest-neighbor hopping, and $t_h$ is the
direct hopping between vertically aligned sites. Its single-layer
counterpart is a standard lattice model for superconductivity across the
BCS--BEC crossover~\cite{Micnas1990,Toschi2005}, whose pairing and
charge correlations have also been studied by quantum Monte
Carlo~\cite{Scalettar1989,Fontenele2022}. The basic ingredients of
\eqn{eqn:H}, including honeycomb optical lattices~\cite{Tarruell2012},
bilayer Hubbard systems~\cite{Gall2021BilayerHubbard}, and attractive
Fermi--Hubbard gases~\cite{Gall2020}, have been realized separately in
ultracold-atom platforms~\cite{BlochDalibardZwerger2008,Paiva2010}.
Bilayer honeycomb systems have also been studied in connection with
valley and non-Abelian charge responses~\cite{Ghadimi2024}, while the
interplay of disorder and interactions in the same bilayer geometry has
been investigated by determinant quantum Monte Carlo~\cite{Prasad2024}.

The interaction is written in particle-hole-symmetric form, so that
$\mu=0$ is exactly half filling and no implicit Hartree shift is
present. At this point the bipartite lattice has the $\eta$-pairing
pseudospin SU(2) symmetry~\cite{Yang1989,Zhang1990}. The uniform
$s$-wave paired state and the bilayer-staggered charge-density-wave
state are therefore exactly degenerate members of the same pseudospin
multiplet [\fig{fig:sectors}(a); Sec.~S3~\cite{SM}]. We choose the
paired representative of this manifold as the saddle about which the
Gaussian fluctuations are constructed.

In the four-sublattice basis $(A_1,B_1,A_2,B_2)$ the normal-state Bloch
Hamiltonian $h(\mathbf k)$ contains the honeycomb structure factor
$f(\mathbf k)=-\sum_{\bm\delta}e^{i\mathbf k\cdot\bm\delta}$,
together with vertical matrix elements $-t_h$ between $A_1$--$A_2$ and
$B_1$--$B_2$, and $-\mu$ on the diagonal (Sec.~S8~\cite{SM}). We set
$t=1$ below. Because the two layers are identical, the layer degree of
freedom diagonalizes exactly into bonding ($\sigma=-$) and antibonding
($\sigma=+$) sectors, each a two-band honeycomb problem, with
eigenvalues
\begin{equation}
  \varepsilon_{\sigma,s}(\mathbf k) = \sigma\,t_h + s\,t\,|f(\mathbf k)| - \mu,
  \qquad \sigma,s=\pm1,
  \label{eqn:bands}
\end{equation}
verified numerically against the full matrix across the Brillouin zone.
Thus every single-layer Bloch band is reproduced in the two parity
sectors with a rigid separation $2t_h$. This momentum-independent
splitting is the microscopic structure that underlies the
relative-phase result derived below.

We take the paired saddle to be uniform in layer and sublattice space,
with $\Delta_0=|U|\langle c_{il\dn}c_{il\up}\rangle$. In the parameter
range studied, this in-phase solution has lower BdG condensation energy
than the layer-antisymmetric configuration $\Delta_1=-\Delta_2$. At half
filling, $\mu=0$ by particle-hole symmetry, and the self-consistency
condition becomes
\begin{equation}
  \frac{4}{|U|} = \frac{1}{N_k}\sum_{\mathbf k}\left[
    \frac{\tanh(\beta E_+/2)}{E_+} + \frac{\tanh(\beta E_-/2)}{E_-}
  \right],
  \label{eqn:gapeq}
\end{equation}
with $E_\pm(\mathbf k)=\sqrt{(|f(\mathbf k)|\pm t_h)^2+\Delta_0^2}$, each
doubly spin degenerate. At the reference point $|U|=4t$ and
$t_h=0.6t$, a $500\times500$ momentum mesh gives
$\Delta_0(T{\to}0)=1.355\,t$ and $T_c^{\rm MF}=0.742\,t$,
corresponding to $2\Delta_0/T_c^{\rm MF}=3.651$, close to the
weak-coupling BCS value. All finite-temperature statements below refer
to this self-consistent Gaussian saddle; $T_c^{\rm MF}$ is the scale at
which it ceases to exist, not a true finite-temperature ordering
transition of the strictly two-dimensional half-filled
SU(2)-symmetric model.

Interlayer hybridization qualitatively changes the pairing threshold.
At half filling, for $0<t_h<3t$, the bonding and antibonding bands
overlap at the Fermi level, producing a finite Fermi surface and hence $U_c=0$ at mean-field
level. This contrasts with the isolated honeycomb layer, whose Dirac
density of states vanishes at the Fermi energy and therefore requires a
finite attraction to pair. Setting $t_h=0$, the same gap equation gives
$U_c^{\rm MF}/t=2.232$, in agreement with the known mean-field
result~\cite{ZhaoParamekanti2006}, while quantum Monte Carlo places the
interacting critical coupling substantially
higher~\cite{Bouadim2009,Otsuka2016,Kennedy2025} (Sec.~S2~\cite{SM}).

\section{Gaussian fluctuations: Goldstone mode and amplitude
threshold}
\label{sec:gaussian}

Expanding the pair field to quadratic order about the uniform saddle,
$\Delta(\mathbf r,\tau)=\Delta_0+\eta_A+i\Delta_0\eta_P$, and
integrating out the fermions gives the Gaussian amplitude and phase
fluctuations. At half filling and $\mathbf q=0$, particle-hole symmetry
decouples these two sectors. In the layer-symmetric, sublattice-uniform
channel the inverse fluctuation propagators take the simple forms
$M_{PP}=-\omega^2I(\omega)$ and
$M_{AA}=(4\Delta_0^2-\omega^2)I(\omega)$, with
$I(\omega)=N_k^{-1}\sum_{\mathbf k,s}\tanh(\beta E_s/2)
/[2E_s(4E_s^2-\omega^2)]$, after the gap equation eliminates $1/|U|$.

The phase kernel therefore vanishes at $\omega=0$, $M_{PP}(0)=0$, as
required by the Anderson--Bogoliubov Goldstone mode associated with
spontaneous U(1) symmetry
breaking~\cite{Anderson1958,Bogoliubov1958}. This zero-frequency
condition is the Goldstone--Ward anchor that will be central below. The
amplitude kernel instead vanishes at $2\Delta_0$, giving the familiar
Higgs threshold~\cite{Pekker2015,Shimano2020,TsujiAoki2015}, whose
resonant excitation has been studied in terahertz experiments on
conventional
superconductors~\cite{MatsunagaShimano2014,Giorgianni2019}. Unlike the
Goldstone mode, however, this amplitude feature is not an isolated
collective pole: $2\Delta_0$ is simultaneously the onset of the
\emph{intra}-parity two-quasiparticle continuum, where $I(\omega)$
becomes nonanalytic.

The layer-odd relative-phase sector considered in \sect{sec:leggett}
differs precisely in this respect. Its vertex connects bonding and
antibonding parity partners, so the relevant continuum is
\emph{inter}-parity rather than the usual intra-parity threshold
$2\Delta_0$. Its onset is
$\omega_c^-=\min_{\mathbf k}(E_++E_-)$, and the general inequality
proved in \eqn{eqn:contbound} gives $\omega_c^->2t_h$ whenever
$\Delta_0>0$. For the half-filled honeycomb lattice,
$E_\pm=[(|f(\mathbf k)|\pm t_h)^2+\Delta_0^2]^{1/2}$. Since $E_++E_-$ is
convex in $|f|$ and has vanishing derivative at $|f|=0$, its minimum
occurs at the Dirac points, giving
$\omega_c^-=2\sqrt{t_h^2+\Delta_0^2}=2.963\,t$ at the reference point.
Thus $2t_h=1.2\,t$ lies well below both the inter-parity continuum and
the conventional amplitude threshold $2\Delta_0=2.709\,t$.

For later comparison of the symmetric and antisymmetric phase sectors,
we use the dimensionless kernels
$K_{\pm P}(\omega)=1-g_0\chi^{\rm raw}_{\pm P}(\omega)$, with
$g_0=1/\chi^{\rm raw}_{+P}(0)=-|U|/8$ fixed by the gap equation. The
Goldstone condition is therefore $K_{+P}(0)=0$. The same vertex $g_0$
enters the layer-even and layer-odd phase channels because the on-site
attraction is identical in the two layers, so their symmetric and
antisymmetric combinations inherit the same interaction strength. This
common Ward-fixed normalization is what allows the zero-frequency phase
anchor to be related to the finite-frequency layer-odd kernel in
\sect{sec:leggett}.

\section{The layer-odd relative-phase kernel: an exact zero at $2t_h$}
\label{sec:leggett}

\begin{figure}[t]
  \centerline{\includegraphics[width=\linewidth]{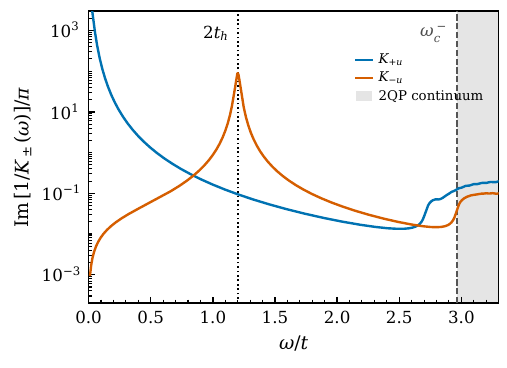}}
  \caption{The two phase channels. Gaussian phase-propagator intensity
  ${\rm Im}[1/K_{\pm u}(\omega)]/\pi$ at $|U|=4t$ and $t_h=0.6t$ on a
  logarithmic scale; $1/K$ is an auxiliary-field propagator, to be
  distinguished from the physical response of \fig{fig:observable}. The layer-symmetric channel $K_{+u}$ carries the
  Anderson--Bogoliubov Goldstone pole as $\omega\to0$. At half filling,
  where the copy-odd amplitude decouples, the sublattice-uniform
  layer-odd channel $K_{-u}$ has an isolated relative-phase pole at
  $\omega=2t_h$ (dotted). The pole lies strictly below the
  inter-parity two-quasiparticle continuum coupled to this channel,
  whose onset is $\omega_c^-=2.963\,t$ (dashed; shaded continuum). The
  separation $\omega_c^->2t_h$ follows generally from
  \eqn{eqn:contbound} whenever $\Delta_0>0$.}
  \label{fig:spectra}
\end{figure}

Beyond the layer-symmetric Goldstone and amplitude sectors, the bilayer
supports layer-\emph{odd} fluctuations generated by
$\lambda_z\otimes\{\tau_x,\tau_y\}$, in contrast to the layer-even
vertices $\lambda_0\otimes\{\tau_x,\tau_y\}$ ($\lambda_a$ acting in
layer space and $\tau_a$ in Nambu space) [\fig{fig:spectra}].
Interlayer hopping removes the independent U(1) phase freedom of the
two layers, so their relative phase is not a Goldstone degree of
freedom: hybridization provides a finite restoring scale for
oscillations of one condensate phase against the other. The channel
relevant to this motion is the sublattice-uniform layer-odd phase
channel $P_{-u}$, whose vertex is the identity in sublattice space. It
is the bilayer Leggett-type mode considered throughout this work.
Because the four-site unit cell also supports sublattice-staggered
internal phase fluctuations, the layer-odd sector contains additional
lattice-specific collective channels; their frequencies are not fixed
by the interlayer splitting and their complete inventory is given in
Sec.~S9~\cite{SM}. The result below concerns only $P_{-u}$, for which
the rigid bonding--antibonding splitting of \eqn{eqn:splitting}
produces an exact Gaussian kernel zero.

The key microscopic ingredient is the rigid bonding--antibonding
splitting,
\begin{equation}
  \xi_{A,s}(\mathbf k,\mu) - \xi_{B,s}(\mathbf k,\mu) = 2t_h,
  \label{eqn:splitting}
\end{equation}
which is independent of momentum, band index, and chemical potential
($A,B$ label the two parity sectors, not the sublattices). The
layer-odd phase vertex is odd under layer parity but diagonal in the
band index, and therefore couples the bonding and antibonding partners
of the same Bloch band. The corresponding two-quasiparticle excitation
has energy $E_A+E_B$. Its phase coherence factor is
$w_{AB}=\mathcal N^{\rm ph}_{AB}/(2E_AE_B)$, with the standard case-II
BCS numerator~\cite{Tinkham}
$\mathcal N^{\rm ph}_{AB}=E_AE_B+\xi_A\xi_B+\Delta_0^2$. The same rigid
splitting in \eqn{eqn:splitting} gives the on-shell identity
$(E_A+E_B)^2-(2t_h)^2=2\mathcal N^{\rm ph}_{AB}$. Thus, precisely at
$\omega=2t_h$, the coherence-factor numerator
cancels the corresponding energy denominator in the layer-odd bubble:
\begin{equation}
\begin{split}
  \chi^{\rm raw}_{-P}(2t_h,\mathbf k)
  &=2\,\frac{\mathcal N^{\rm ph}_{AB}}{2E_AE_B}\,
   \frac{2(E_A{+}E_B)}{-2\mathcal N^{\rm ph}_{AB}}
  =-\Big(\frac{1}{E_A}+\frac{1}{E_B}\Big)\\[2pt]
  &=\chi^{\rm raw}_{+P}(0,\mathbf k),
\end{split}
  \label{eqn:pointwise}
\end{equation}
pointwise in momentum and band. Eigenvector conventions, normalization
factors, and independent numerical checks are given in
Sec.~S10~\cite{SM}.

Equation~(\ref{eqn:pointwise}) is the central cancellation. The
symmetric phase bubble at zero frequency is fixed by the
Goldstone--Ward condition $K_{+P}(0)=0$, while the symmetric and
antisymmetric pair fields inherit the same local interaction vertex.
\eqn{eqn:pointwise} therefore transfers that zero-frequency condition
directly to the layer-odd channel, $K_{-P}(2t_h)=0$. The result does
not depend on the magnitude of $|U|$ within the paired Gaussian theory.
Only the rigid splitting $\xi_A-\xi_B=2t_h$ and the common gap
$\Delta_0$ enter the cancellation.

The physical content can be stated in terms of three structures. Global
U(1) symmetry and self-consistency fix the common-phase channel at zero
frequency through the Goldstone--Ward identity. Interlayer
hybridization produces a momentum-independent parity splitting $2t_h$.
In copy space the same hybridization acts as a uniform pseudomagnetic
field, giving the exact kinetic eigenoperator relation
$[H_{\rm kin},L]=-2t_hL$ for $L=T_z-iT_y$ (\sect{sec:larmor}). The
rigid splitting therefore shifts the Ward-anchored phase structure from
zero frequency in the symmetric channel to $2t_h$ in the copy-odd
channel. \eqn{eqn:pointwise} is the momentum-resolved algebraic
expression of this Ward--Larmor mechanism.

The Larmor relation should not be read as a conservation law. $L$ is an
eigenoperator of $H_{\rm kin}$, not a conserved quantity, and the
conservation content enters through the U(1) Ward identity.
Accordingly, $K_{-P}(2t_h)=0$ is an exact statement about the Gaussian
pairing-field kernel rather than a separate Ward identity of the full
interacting Hamiltonian. The mechanism is sketched in
\fig{fig:mechanism}. At particle-hole symmetry, where the odd
amplitude decouples, this kernel zero becomes the isolated
relative-phase pole shown in \fig{fig:spectra}.

Two different pseudospin structures enter the pinning mechanism. In
Nambu space, each bonding or antibonding parity sector is described by
an Anderson pseudospin in the effective field $(\Delta_0,0,\xi)$.
Superconducting self-consistency makes a uniform transverse rotation of
these pseudospins costless, giving the Goldstone--Ward anchor
$K_{+P}(0)=0$. The finite scale $2t_h$, however, belongs to copy space
rather than Nambu space. Interlayer hopping splits the bonding and
antibonding copies of the same BdG problem by the rigid amount $2t_h$,
and the layer-odd phase vertex connects precisely these parity
partners. Consequently, the coherence-factor structure of the
symmetric phase channel at $\omega=0$ is reproduced in the
antisymmetric channel at $\omega=2t_h$; \eqn{eqn:pointwise} is the
pointwise statement of this correspondence. The pinned scale is
therefore distinct from the ordinary BdG pseudospin-precession scale:
at the reference point the minimum $2E_{\mathbf k}$ is
$2\Delta_0=2.26\times2t_h$.

\subsection{The general two-copy theorem}
\label{sec:general}

The derivation of \eqn{eqn:pointwise} does not rely on the honeycomb
dispersion. Its essential ingredients are instead a rigid splitting
between two identical copies and a common pairing structure that treats
those copies equivalently. Consider
\begin{equation}
  H_0(\mathbf k) \;=\; h(\mathbf k)\otimes\lambda_0 \;-\; t_h\,\lambda_x,
  \label{eqn:twocopy}
\end{equation}
where $h(\mathbf k)$ is a Hermitian Bloch Hamiltonian with an arbitrary
number of orbitals and in any spatial dimension, while $\lambda_a$ acts
in copy space. Because the intercopy hybridization $t_h$ is
momentum independent and proportional to the identity in the internal
orbital space, rotating to bonding and antibonding copy parity gives
the two blocks $h(\mathbf k)\mp t_h$. Every band of $h(\mathbf k)$
therefore acquires a parity partner separated by the same $2t_h$, so
$\xi_{A,n}-\xi_{B,n}=2t_h$ for every band $n$, momentum $\mathbf k$,
and chemical potential.

The pairing structure supplies the second ingredient. We assume
identical copies, a momentum-independent $t_h$, no copy-antisymmetric
one-body potential, and a common saddle
$\hat\Delta=\Delta_0\,\openone_{\rm orb}\otimes\lambda_0$. For a
multiorbital system we additionally require conventional singlet
pairing between time-reversed partners to remain band diagonal,
equivalently $h^{\mathsf T}(-\mathbf k)=h(\mathbf k)$ in an appropriate
pairing basis, or its antiunitary equivalent. Under these conditions
each parity sector reduces band by band to the usual BdG form
$H_\sigma=\xi_\sigma\tau_z+\Delta_0\tau_x$. Outside this class, the
particle and hole blocks need not be diagonalized in the same band
basis, the common gap can generate interband pairing, and the reduction
underlying \eqn{eqn:pointwise} no longer follows
(Secs.~S1 and S10~\cite{SM}).

Within this class, the band dispersion itself drops out of the pinning
argument. For every parity pair, both the phase coherence factor and
the on-shell denominator depend only on $(\xi_A,\xi_B,\Delta_0)$, with
the sole constraint $\xi_A-\xi_B=2t_h$. The same pointwise cancellation
that produced \eqn{eqn:pointwise} therefore applies band by band,
giving $K_{-P}(2t_h)=0$ for any $h(\mathbf k)$, orbital number,
dimensionality, and chemical potential satisfying the conditions above.

The same structure also determines whether the $2t_h$ kernel zero is
separated from quasiparticle excitations, without reference to any
particular band dispersion. The layer-odd uniform phase vertex connects
opposite parity partners, so its pair-creation continuum begins at
$\omega_c^-=\min_{\mathbf k}[E_A(\mathbf k)+E_B(\mathbf k)]$. Pointwise
in momentum,
\begin{equation}
\begin{split}
  E_A+E_B &=\sqrt{\xi_A^2+\Delta_0^2}+\sqrt{\xi_B^2+\Delta_0^2}\\
  &>|\xi_A|+|\xi_B|\ \ge\ |\xi_A-\xi_B|=2t_h ,
\end{split}
  \label{eqn:contbound}
\end{equation}
where the first inequality is strict for $\Delta_0>0$ and the second is
the triangle inequality. Hence $\omega_c^->2t_h$ throughout the paired
state for every member of the class \eqn{eqn:twocopy}. The Gaussian
kernel zero at $2t_h$ therefore lies strictly below the inter-parity
pair-creation continuum and cannot be a threshold singularity. At
particle-hole symmetry, where the odd amplitude decouples, this
immediately makes it an isolated undamped relative-phase pole.

Finite temperature does not close this protection while the paired
saddle remains gapped. Thermal occupation introduces a
quasiparticle-scattering branch at $\Omega=|E_A-E_B|$, but
$E(\xi)=\sqrt{\xi^2+\Delta_0^2}$ is strictly contractive for
$\Delta_0>0$, since $|dE/d\xi|=|\xi|/E<1$. Therefore
$|E_A-E_B|<|\xi_A-\xi_B|=2t_h$, while \eqn{eqn:contbound} keeps pair
creation above $2t_h$. The pinned frequency thus lies in a two-sided
spectral window, $|E_A-E_B|<2t_h<E_A+E_B$, with thermal scattering
below and pair creation above. The
finite-temperature kernel identity remains exact because the two
branches separately cancel their on-shell coherence factors and
recombine into the symmetric-channel Ward value (Sec.~S10~\cite{SM}).

It is important to distinguish this continuum isolation from the
position of the full collective pole. The diagonal identity
$K_{-P}(2t_h)=0$ holds at arbitrary chemical potential, but the
physical eigenfrequency is determined by the coupled amplitude--phase
determinant. Particle-hole symmetry makes the off-diagonal element
$K^-_{AP}$ vanish and pins the pole to $2t_h$; away from that symmetry,
$K^-_{AP}\neq0$ and the collective frequency shifts, with
$\Omega_{\rm full}-2t_h=O(\delta n^2)$ near half filling
(\sect{sec:robust}). For the same reason, comparing $2t_h$ with the
conventional pair-breaking scale $2\Delta_0$ can be misleading. For
example, at $t_h=1.5t$ and $\Delta_0=0.4t$, one has
$2t_h=3.0t>2\Delta_0=0.8t$, yet the relevant inter-parity threshold
remains $\omega_c^-=3.105\,t>2t_h$. The $2t_h$ kernel zero therefore
remains below the inter-parity continuum to which the uniform odd phase
channel couples.

The same assumptions also identify where the $2t_h$ theorem ceases to
apply. A momentum-dependent hybridization $t_h(\mathbf k)$ removes the
single rigid parity splitting required for the pointwise mapping of the
Ward anchor. Unequal copies or unequal gaps destroy the equivalent,
decoupled parity-pair BdG structure, so the relative-phase frequency is
no longer fixed by $2t_h$ and becomes model dependent, generally
involving both hybridization and relative-phase (Josephson)
stiffness~\cite{Leggett1966}. A copy-antisymmetric one-body potential
$v\lambda_z$ instead tilts the copy pseudofield away from the
hybridization axis: the normal-state splitting becomes
$2\sqrt{t_h^2+v^2}$, and the original layer-odd vertex is no longer
purely transverse to that field. Thus identical copies, a
momentum-independent scalar $t_h$, a common copy-symmetric gap, no
copy-antisymmetric one-body field, and conventional bandwise singlet
pairing constitute sufficient conditions for the exact Gaussian kernel
zero $K_{-P}(2t_h)=0$. They are not claimed to be necessary
conditions.

\subsection{Operator (Larmor) form of the theorem}
\label{sec:larmor}

\begin{figure}[t]
  \centerline{\includegraphics[width=\linewidth]{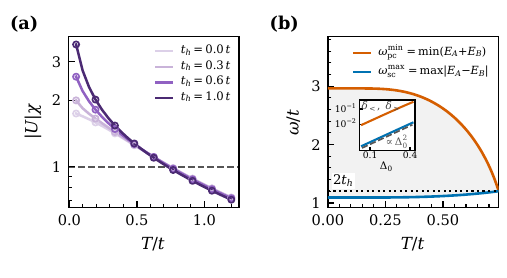}}
  \caption{Symmetry and spectral structure at the
  particle-hole-symmetric saddle. (a) Pairing and bilayer-staggered
  charge-order susceptibilities at half filling. For every $t_h$, the
  two coincide and reach the Thouless criterion $|U|\chi=1$ at the same
  temperature, reflecting the exact pseudospin equivalence of the
  pairing and charge-order channels; the paired saddle used here is one
  representative of this manifold. (b) Two-sided spectral window at
  $|U|=4t$ and $t_h=0.6t$. Throughout the paired state,
  $\Delta_0(T)>0$, the thermal-scattering edge
  $\omega_{\rm sc}^{\max}=\max_{\mathbf k,n}|E_A-E_B|$ remains strictly
  below $2t_h$ (dotted), while the pair-creation onset
  $\omega_{\rm pc}^{\min}=\min_{\mathbf k,n}(E_A+E_B)$ remains strictly
  above it. Inset: for the half-filled honeycomb spectrum, the lower
  and upper separations from $2t_h$ both vanish as $\Delta_0^2$ on
  approaching the mean-field transition (fitted exponents $1.99$ and
  $1.96$). This closing is specific to the honeycomb realization and is
  not required by the general inequalities (Sec.~S10~\cite{SM}).}
  \label{fig:sectors}
\end{figure}

Interlayer hopping has a simple operator interpretation: it acts as a
uniform pseudomagnetic field in layer space and sets the natural
precession scale $2t_h$. This is the physical origin of the rigid
bonding--antibonding splitting in \eqn{eqn:splitting} and of the finite
frequency that enters the Gaussian kernel theorem. Let
$T_a=\tfrac12\sum_{i\sigma}c^\dagger_{il\sigma}(\lambda_a)_{ll'}
c_{il'\sigma}$ denote the layer pseudospin, with layer indices summed. The interlayer
hopping is exactly
\begin{equation}
  H_\perp = -2t_h\,T_x ,
  \label{eqn:field}
\end{equation}
while the identical intralayer Hamiltonians are proportional to the
identity in layer space and therefore commute with every $T_a$. The
transverse combination $L\equiv T_z-iT_y$ is consequently an
eigenoperator of the kinetic Hamiltonian,
\begin{equation}
  [H_{\rm kin},\,L] = -2t_h\,L .
  \label{eqn:larmor}
\end{equation}
Thus the layer pseudospin precesses about the hybridization field at
the Larmor frequency $2t_h$, independent of the intralayer dispersion,
momentum, or chemical potential [\fig{fig:mechanism}, right]. The rigid
parity splitting $\xi_A-\xi_B=2t_h$ appearing in \eqn{eqn:splitting} is
the single-particle spectral manifestation of this same copy-space
Larmor algebra.

Two qualifications are essential. First, \eqn{eqn:larmor} alone does
not pin the relative-phase collective mode. The pairing field does not
commute with $L$, and the rigorous Gaussian result follows only when
the copy-space Larmor scale is combined with the Goldstone--Ward
condition through the pointwise identity \eqn{eqn:pointwise}. Second,
the Larmor precession itself belongs to the particle--hole sector. The
layer-odd \emph{pair} operator
$P^\dagger_-=(P^\dagger_1-P^\dagger_2)/\sqrt2$ satisfies
$[H_\perp,P^\dagger_-]=0$. In the bonding--antibonding basis
$P^\dagger_-$ creates one fermion in each parity sector, so the
opposite hybridization energies cancel and the pair operator itself
carries no $2t_h$ precession. Consistently, $P^\dagger_-$ and $L$
belong to different particle-number sectors and are orthogonal,
${\rm Tr}(P^\dagger_-L^\dagger)=0$. Thus the $2t_h$ zero of the
Gaussian pair kernel should not be interpreted as the eigenfrequency of
$P^\dagger_-$; it arises from the Ward--Larmor mapping embodied in
\eqn{eqn:pointwise}.

A third pseudospin structure is present at half filling but plays a
different role. The $\eta$-pairing pseudospin of \sect{sec:model}
controls the exact degeneracy between the paired and staggered
charge-ordered states, whereas the layer pseudospin $T_a$ describes the
copy-space dynamics generated by interlayer hybridization. The two are
not independent commuting quantum numbers: for the bipartite (Shiba)
form of the $\eta$ generator~\cite{Shiba1972},
$[T_z,\eta^\dagger]\neq0$. The $\eta$-pseudospin therefore organizes
the half-filled order-parameter manifold, while the layer pseudospin
supplies the Larmor scale entering the pinning mechanism.

\subsection{Sector structure and the density channel}
\label{sec:sectors}

At the particle-hole-symmetric saddle, the sublattice-uniform layer-odd
phase fluctuation is dynamically decoupled from the other
order-parameter sectors. Its cross-bubbles with the layer-symmetric
density, the bilayer-staggered charge-density-wave pattern
$(+,-,-,+)$, and both amplitude channels vanish exactly. The
$\eta$-pairing partner of the superconducting state therefore does not
hybridize with the relative-phase oscillation, and the zero of
$K_{-P}$ remains a simple isolated pole; the corresponding kernel
derivative is finite throughout the coupling range studied
(Sec.~S6~\cite{SM}).

The layer-odd density $n_-=N_1-N_2$ is different. Its
cross-susceptibility with the relative phase is finite and purely
imaginary at $\omega=2t_h$. This is the response-function
manifestation of the copy-pseudospin precession of \sect{sec:larmor}:
relative density and relative phase form the conjugate variables of the
layer-odd oscillation. Consequently, a perturbation coupling directly
to $n_-$ can excite and detect the relative-phase mode, as developed in
\sect{sec:observable}.

The finite phase--density coupling does not mean that $n_-$ should be
introduced as an additional independently resummed
Hubbard--Stratonovich field. The microscopic interaction in
\eqn{eqn:H} was decoupled in the pairing channel, and adding an
independent density resummation double counts the same interaction
content. Indeed, such a construction shifts two zero-frequency poles
that are fixed by exact symmetries, including the $t_h\to0$
relative-phase Goldstone limit and the pseudospin-related
staggered-density Goldstone mode (Sec.~S3~\cite{SM}). The consistent
Gaussian theory therefore retains the pairing field as the fluctuating
collective field, while the layer-odd density enters as a physical
probe. Its coupling determines the observable character and spectral
weight of the mode, not the Gaussian kernel frequency.

Finally, the pointwise identity \eqn{eqn:pointwise} and the associated
Ward value were checked independently using the full $8\times8$ BdG
Lehmann representation. They hold to double-precision round-off even on
a $4\times4$ momentum mesh, confirming that the equality is pointwise
rather than a consequence of Brillouin-zone convergence
(Secs.~S4 and S10~\cite{SM}).

\section{Layer-imbalance response and exact sum rule}
\label{sec:observable}

\begin{figure}[t]
  \centerline{\includegraphics[width=\linewidth]{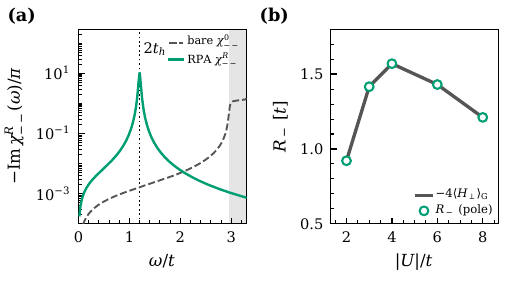}}
  \caption{Layer-imbalance response of a fixed microscopic operator.
  The probe is $H_{\rm probe}(t)=-V_-(t)\,n_-$, and $\chi^R_{--}$ is
  defined per unit cell (four sites, two per layer); half filling
  throughout. (a) Spectral function
  $-{\rm Im}\,\chi^R_{--}(\omega)/\pi$ at $|U|=4t$ and $t_h=0.6t$ on a
  logarithmic scale. The conserving Gaussian response of
  \eqn{eqn:chinn} (vermilion) contains the isolated pole at
  $\omega=2t_h$, while the bare bubble (grey dashed) has no
  subcontinuum pole and begins only at the inter-parity threshold
  $\omega_c^-$ (shaded continuum). The finite width of the $2t_h$ peak
  and any residual subcontinuum tail away from it arise from the
  Lorentzian broadening $\eta$. (b) Pole residue $R_-$ (circles)
  compared with the independently evaluated Gaussian first moment
  $S_1^{\rm G}=-4\langle H_\perp\rangle_{\rm G}$ (line). Their equality
  demonstrates the single-pole exhaustion of \eqn{eqn:exhaust}
  throughout the coupling range shown. The full-interacting moment
  $-4\langle H_\perp\rangle$ is distinct and enters the exact spectrum
  of \fig{fig:ed}. The dimensionless ratio $R_-/S_1^{\rm G}$, unlike
  the convention-dependent pairing-field kernel residue, is invariant
  under a rescaling of the microscopic probe $n_-\to a\,n_-$
  (Sec.~S6~\cite{SM}).}
  \label{fig:observable}
\end{figure}

The finite phase--density coupling identified in \sect{sec:sectors}
provides a direct physical probe of the relative-phase oscillation. We
therefore consider an interlayer bias
$H_{\rm probe}(t)=-V_-(t)\,n_-$, with $n_-=N_1-N_2$, and define the
retarded layer-imbalance response $\chi^R_{--}$ per unit cell. Unlike
the residue of a Hubbard--Stratonovich field, whose magnitude depends
on the normalization chosen for the collective coordinate, $n_-$ is a
fixed microscopic operator. Its susceptibility and spectral weight
therefore have an unambiguous physical normalization.

At particle-hole symmetry the copy-odd amplitude decouples,
$K^-_{AP}=\chi^0_{n_-A_-}=0$, so the layer imbalance couples to the
relative-phase sector alone. Within the same pairing-field Gaussian
theory in which the pinning theorem holds, the dressed response is
\begin{equation}
  \chi^R_{--}(\omega)=\chi^{0}_{--}(\omega)
  +g_0\,\frac{\chi^{0}_{n_-\phi_-}(\omega)\,\chi^{0}_{\phi_-n_-}(\omega)}
             {K_{-P}(\omega)} .
  \label{eqn:chinn}
\end{equation}
Because $K_{-P}(2t_h)=0$ while the phase--density cross-susceptibility
is finite there, $\chi^R_{--}$ contains an isolated pole at
$\Omega_L=2t_h$ [\fig{fig:observable}(a)]. Thus the copy-space Larmor
dynamics is visible in the response of the microscopic layer imbalance
rather than only in an auxiliary pairing-field propagator.

The total energy-weighted spectral weight obeys a much stronger
constraint that is independent of the Gaussian approximation. For the
full interacting Hamiltonian, the intralayer hopping and
chemical-potential terms conserve $N_1$ and $N_2$ separately, while the
on-site interaction is diagonal in the layer occupations. They
therefore commute with $n_-$, leaving only the interlayer hopping in
the double commutator. Consequently,
\begin{equation}
\begin{split}
  S_1&\equiv\int_0^\infty\!\!d\omega\,2\omega
  \left[-\tfrac{1}{\pi}{\rm Im}\,\chi^R_{--}(\omega)\right]\\
  &=\langle[n_-,[H,n_-]]\rangle=-4\langle H_\perp\rangle .
\end{split}
  \label{eqn:sumrule}
\end{equation}
\eqn{eqn:sumrule} is an exact operator sum rule at arbitrary
interaction strength. Physically, it says that although interactions
can redistribute the layer-imbalance spectrum and shift its
characteristic frequencies, its first moment is controlled entirely by
the interlayer kinetic energy. What is special about the Gaussian
theory is not the sum rule itself, but the stronger result shown below:
at particle-hole symmetry the single pole at $2t_h$ exhausts it
completely.

At particle-hole symmetry, the Gaussian response obeys a stronger
result than the exact sum rule alone: the entire first moment is
carried by the $2t_h$ pole. Using the kernel convention
$K=1-g_0\chi^{\rm raw}$ of \sect{sec:gaussian}, define $d\equiv2t_h$,
$s\equiv E_A+E_B$, and $p\equiv E_AE_B$. The coherence-factor
identities of Secs.~S11 and S12~\cite{SM}, together with the
Ward/gap-equation value $g_0=1/\chi^{\rm raw}_{+P}(0)$, reduce the
kernel, phase--density cross bubble, and bare density bubble in
\eqn{eqn:chinn} to the same parity-pair sums. Their continuum
contributions then cancel identically, leaving
\begin{equation}
  \chi^R_{--}(\omega)=\frac{4\Delta_0^2d^2\,G}{\omega^2-d^2}
  =\frac{-4\langle H_\perp\rangle_{\rm G}}{\omega^2-d^2}
  =\frac{S_1^{\rm G}}{\omega^2-d^2} ,
  \label{eqn:exhaust}
\end{equation}
with $G=N_k^{-1}\sum_{\mathbf k,n}s/[p(s^2-d^2)]$ and
$S_1^{\rm G}\equiv-4\langle H_\perp\rangle_{\rm G}$. Thus the physical
pole residue is $R_-=4\Delta_0^2d^2G=S_1^{\rm G}$: within the
particle-hole-symmetric Gaussian theory, the vertex correction removes
the bare continuum contribution and the single pole at $2t_h$ exhausts
the layer-imbalance first moment.

This exhaustion and the sum rule itself have different status.
\eqn{eqn:sumrule}, $S_1=-4\langle H_\perp\rangle$, is an exact operator
identity of the full interacting Hamiltonian. \eqn{eqn:exhaust} is
stronger but Gaussian: it states that the corresponding conserving
Gaussian response carries its moment
$S_1^{\rm G}=-4\langle H_\perp\rangle_{\rm G}$ entirely in one pole.
The saddle expectation $\langle H_\perp\rangle_{\rm G}$ is generally
different from the exact interacting expectation
$\langle H_\perp\rangle$, with the two coinciding at $U=0$.
Interactions must therefore continue to satisfy the exact many-body
moment $S_1=-4\langle H_\perp\rangle$, but they are free to shift the
pole and redistribute that moment among several excitations.
\fig{fig:observable}(b) verifies the Gaussian exhaustion
$R_-=S_1^{\rm G}$, while \sect{sec:beyond} and \fig{fig:ed} show how
the exact interacting spectrum departs from it.

The algebra behind $R_-=S_1^{\rm G}$, like the cancellation in
\eqn{eqn:pointwise}, is pointwise in the parity-pair variables
$(\xi_A,\xi_B,\Delta_0)$ and does not depend on the honeycomb
dispersion. Its realization as the complete physical layer-imbalance
response, however, requires particle-hole symmetry, because only then
do the odd amplitude and phase sectors decouple (Sec.~S12~\cite{SM}).
This distinction is important: the underlying algebra is general within
the two-copy class, whereas single-pole exhaustion is not.

The microscopic reason that $n_-$ couples so directly to the mode is
again the layer-pseudospin structure. With $L=T_z-iT_y$ and
$n_-=2T_z$,
\begin{equation}
  n_- = L + L^\dagger ,
  \label{eqn:nLL}
\end{equation}
so the layer imbalance is the transverse copy-pseudospin coordinate
together with its conjugate. Away from particle-hole symmetry,
$K^-_{AP}$ and $\chi^0_{n_-A_-}$ become nonzero, the density response
also probes the odd-amplitude sector, and the physical pole is
determined by the coupled amplitude--phase determinant rather than by
$K_{-P}(2t_h)=0$ alone. The diagonal phase-kernel zero survives, but
neither the pole at $2t_h$ nor the single-pole exhaustion of
\eqn{eqn:exhaust} is then implied.

A useful response-level corollary follows. At particle-hole symmetry
the phase--density cross bubble vanishes at $\omega=0$, so the Gaussian
vertex correction leaves the static layer-imbalance susceptibility
unchanged. Because the same response is exhausted by a single pole, its
static susceptibility and first moment determine one frequency,
\begin{equation}
  \Omega_L^2=\frac{S_1^{\rm G}}{-\chi^R_{--}(0)}
  =\frac{2t_h\langle T_x\rangle_{\rm G}}{\chi^{\rm stat}_{zz}} ,
  \label{eqn:static}
\end{equation}
with the positive static susceptibility
$\chi^{\rm stat}_{zz}\equiv-\chi^R_{T_zT_z}(0)$, so that
$-\chi^R_{--}(0)=4\chi^{\rm stat}_{zz}$ by $n_-=2T_z$.
Thus, within the particle-hole-symmetric Gaussian theory, the dynamical
pole frequency can be obtained entirely from equilibrium quantities,
without resolving the dynamical spectrum.

The same construction remains meaningful beyond Gaussian order, but its
interpretation changes. Using the exact first moment
$S_1=-4\langle H_\perp\rangle$ and the true static susceptibility
defines the spectral-moment scale
$\Omega_{\rm mom}^2\equiv S_1/[-\chi^R_{--}(0)]$. For a single-mode
response $\Omega_{\rm mom}=\Omega_L$, whereas spectral redistribution
causes the two scales to differ. Comparing the dominant dynamical
excitation with $\Omega_{\rm mom}$ therefore directly measures the
breakdown of Gaussian single-mode exhaustion.

Interaction effects beyond Gaussian order are free to shift the pole
and redistribute its oscillator strength because the interaction torque
destroys the Larmor closure. These two effects need not occur together:
an excitation can move substantially in frequency while remaining
spectrally dominant. The exact statement that survives is the sum-rule
\emph{identity} $S_1=-4\langle H_\perp\rangle$, not a fixed numerical
value of $S_1$. The dimensionless Gaussian residue $R_-/S_1^{\rm G}$
and the bare-bubble sum-rule excess are discussed in
Sec.~S6~\cite{SM}. The equality of the dynamical and moment scales
therefore provides a diagnostic of single-mode exhaustion. Before
turning in \sect{sec:beyond} to its failure under the full interaction,
we first delimit the Gaussian result away from particle-hole symmetry
and at finite momentum.

\section{Away from particle-hole symmetry and at finite momentum}
\label{sec:robust}

The rigid parity splitting of \eqn{eqn:splitting} is independent of
chemical potential, so doping does not alter the pointwise identity
underlying $K_{-P}(2t_h)=0$. What doping changes is the relation between this
diagonal kernel zero and the physical collective pole. Particle-hole
symmetry forces the copy-odd amplitude and phase sectors to decouple.
For a small density detuning $\delta n$ from the
particle-hole-symmetric saddle, the off-diagonal element switches on
linearly, $K^-_{AP}=O(\delta n)$. Since it enters the coupled
amplitude--phase determinant through $K^-_{AP}K^-_{PA}$, the leading
displacement of the collective eigenfrequency is quadratic,
$\Omega_{\rm full}-2t_h=O(\delta n^2)$. Thus the exact $2t_h$ zero of
the diagonal phase kernel survives doping, while the physical
relative-phase pole is pinned to it only at particle-hole symmetry.

Finite momentum removes a different ingredient of the exact
$\mathbf q=0$ construction. The fluctuation now connects states at
$\mathbf k$ and $\mathbf k+\mathbf q$, so the pointwise cancellation
underlying \eqn{eqn:pointwise} no longer fixes the frequency exactly.
Analyticity about $(\omega,\mathbf q)=(2t_h,\mathbf 0)$, together with
the $C_{6v}$ symmetry of the honeycomb lattice, nevertheless constrains
the small-$q$ dispersion at half filling to
$\Omega_L^2(\mathbf q)=(2t_h)^2+\alpha_L|\mathbf q|^2+O(q^4)$, which is
isotropic to quadratic order. We do not quote a numerical value of
$\alpha_L$: finite-difference estimates along $\Gamma\!-\!M$ and
$\Gamma\!-\!K$ differ by approximately $12\%$, indicating that the
available momentum resolution is insufficient for a controlled
extraction. The coupled determinant, the small-doping behavior, and
the finite-$q$ analysis are given in Sec.~S5~\cite{SM}.

These are departures \emph{within} the Gaussian theory: particle-hole
asymmetry separates the diagonal kernel zero from the coupled pole,
while finite momentum removes the pointwise $\mathbf q=0$ cancellation.
A different question is whether the full interaction preserves the
$\mathbf q=0$, particle-hole-symmetric pinning itself. We address that
next.

\section{Beyond Gaussian order: torque, exact rung, and exact diagonalization}
\label{sec:beyond}

It remains to ask whether the Gaussian relative-phase pole at
$\Omega_L=2t_h$ is protected once the Hubbard interaction is treated
beyond Gaussian order. The copy-pseudospin formulation of
\sect{sec:larmor} turns this into an operator question. The kinetic
Hamiltonian closes exactly on the transverse pseudospin,
$[H_{\rm kin},L]=-2t_hL$. If the interaction commuted with $L$, the
same Larmor algebra would survive at the bare frequency $2t_h$; if its
commutator closed back onto $L$, the precession could remain an exact
eigenoperator relation with an interaction-renormalized frequency.
Neither occurs for the Hubbard interaction.

A direct evaluation gives the exact interaction torque
\begin{equation}
  R \equiv [H_U,\,L]
  = \frac{|U|}{2}\sum_{j\sigma}
  \big(n_{j1\bar\sigma}-n_{j2\bar\sigma}\big)
  \big(c^\dagger_{j1\sigma}c_{j2\sigma}+c^\dagger_{j2\sigma}c_{j1\sigma}\big).
  \label{eqn:torque}
\end{equation}
Unlike the one-body Larmor term, $R$ couples interlayer transfer of one
spin species to the instantaneous layer imbalance of the opposite spin.
The copy-pseudospin motion is therefore no longer a closed one-body
precession: interactions couple it directly to two-particle
correlations. Correspondingly, $R$ is nonzero and has no component
along $L$ or the other layer-pseudospin generators, so it cannot be
absorbed into a renormalization of the pseudomagnetic field. The
closed Larmor algebra is therefore lost once the interaction acts
dynamically.

The torque in \eqn{eqn:torque} establishes that the $2t_h$ frequency is
no longer algebraically protected, but it does not determine the
magnitude or even the sign of the resulting shift. At the
particle-hole-symmetric Gaussian saddle the reference behavior is
nevertheless completely fixed: whenever the paired solution has
$\Delta_0(T)>0$, the relative-phase pole remains at $2t_h$ and is
isolated, because the inter-parity pair-creation continuum lies above
$2t_h$ while thermally activated scattering remains below it. An
exactly solvable interacting problem shows what becomes possible once
the interaction is treated without the Gaussian truncation. For a
single vertical rung with the same on-site attraction, the layer-odd
bright excitation has
$\Omega_{\rm rung}=\tfrac12\big(\sqrt{U^2+16t_h^2}-|U|\big)$ at every
coupling (Sec.~S7~\cite{SM}). It reduces to $2t_h$ at $U=0$, decreases
monotonically with $|U|$, and approaches the pair-transfer scale
$4t_h^2/|U|$ at strong coupling. Thus, already in the minimal
interacting problem, the bare hybridization scale evolves into an
interaction-dependent pair-transfer scale. The rung does not determine
the collective response of the lattice, where intralayer motion
provides additional many-body dynamics, so we next treat that dynamics
exactly on a finite periodic cluster.

We diagonalize the full attractive Hubbard Hamiltonian at half filling
on a periodic $N_s=12$ bilayer cluster containing three unit cells per
layer. The cluster preserves all three distinct intralayer neighbours
of every site, $z=3$, and, crucially, retains the $n_-$-bright
excitation at $2t_h$ in the noninteracting limit. The layer-imbalance
spectrum is obtained from the Krylov space generated by $n_-|0\rangle$.
At $U=0$ the response consists of a single bright line at
$\Omega_{\rm dom}=2t_h$ with $f_{\rm dom}=1$, identifying the
excitation followed at finite interaction continuously with the same
microscopic layer-imbalance channel from which the Gaussian $2t_h$ pole
emerges in the paired state. The
exact double-commutator sum rule of \eqn{eqn:sumrule} is satisfied to
better than $8\times10^{-13}$, providing an operator-level check
independent of the location of the individual spectral lines. Cluster
construction and convergence tests are given in Sec.~S13~\cite{SM}.

\begin{figure}[t]
  \centerline{\includegraphics[width=\linewidth]{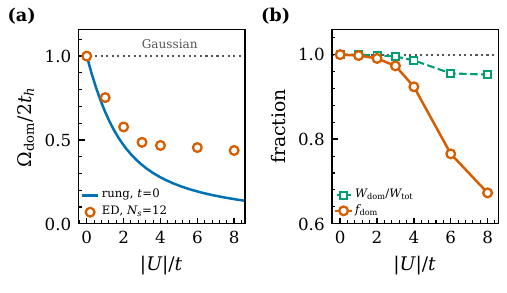}}
  \caption{Exact interacting response beyond Gaussian theory. Exact
  diagonalization of the attractive Hubbard Hamiltonian on a periodic
  $N_s=12$, $z=3$ bilayer cluster at $t_h=0.6t$ and half filling.
  (a) Frequency $\Omega_{\rm dom}$ of the dominant layer-imbalance
  excitation in units of $2t_h$. Gaussian theory gives
  $\Omega_L/(2t_h)=1$ (dotted). The solid curve shows the exactly
  solvable interacting-rung result,
  $\Omega_{\rm rung}/(2t_h)=[\sqrt{U^2+16t_h^2}-|U|]/(4t_h)$, included
  as the $t=0$ limit rather than as a lattice prediction. (b)
  $f_{\rm dom}=2\Omega_{\rm dom}W_{\rm dom}/[-4\langle H_\perp\rangle]$,
  the fraction of the exact first moment in \eqn{eqn:sumrule} carried
  by the dominant line (circles), compared with its zeroth-moment
  spectral-weight fraction $W_{\rm dom}/W_{\rm tot}$ (squares). The
  dominant excitation retains nearly all of the ordinary spectral
  weight while progressively ceasing to exhaust the exact first moment.
  All values are exact for the specified finite cluster; no
  thermodynamic-limit extrapolation is implied.}
  \label{fig:ed}
\end{figure}

Treating the interaction exactly changes both the frequency and the
distribution of oscillator strength [\fig{fig:ed}]. At $|U|=4t$, the
dominant layer-imbalance excitation lies at
$\Omega_{\rm dom}=0.467\,(2t_h)$. It still carries $98.6\%$ of the
zeroth-moment spectral weight, but only $92.4\%$ of the exact first
moment. At $|U|=8t$, the corresponding values are $0.437\,(2t_h)$,
$95.3\%$, and $67.3\%$. The excitation therefore remains spectrally
dominant even after its frequency has been reduced to less than half
the Gaussian value, while a comparatively small amount of spectral weight transferred
to higher energies carries an increasingly large fraction of the
energy-weighted sum rule. This is the many-body departure from the
Gaussian result: the layer-odd excitation survives, but neither its
bare $2t_h$ frequency nor its single-pole exhaustion of the first
moment does.

A finite number-conserving cluster has no spontaneously broken
condensate phase, so $\Omega_{\rm dom}$ is not the thermodynamic
Leggett-mode frequency. It is the exact layer-imbalance excitation of
the finite interacting Hamiltonian, connected continuously to the
$2t_h$ excitation in the same response channel at $U=0$. The
calculation therefore establishes that the Gaussian pinning is not an
exact property of the interacting many-body spectrum, while making no
thermodynamic-limit extrapolation of the renormalized collective
frequency. The exact torque, the interacting rung, and the
finite-cluster spectrum give complementary levels of this conclusion:
the torque removes algebraic protection, the rung demonstrates exact
interaction-driven detuning, and the lattice calculation shows that
strong detuning persists when intralayer dynamics are restored.
Determining the thermodynamic interacting collective-mode frequency and
linewidth requires a symmetry-broken treatment beyond Gaussian order
and lies outside the present work.

\section{Discussion and scope}
\label{sec:disc}

The central result is most naturally understood as a relation between a
microscopic hybridization scale and the collective dynamics of the
paired state. A momentum-independent intercopy hopping acts as a
uniform pseudomagnetic field in copy space. It therefore splits every
bonding--antibonding pair by the same energy $2t_h$, independent of
momentum and of the details of the intralayer dispersion.
Superconductivity supplies a second ingredient. Self-consistency fixes
the common phase fluctuation at zero frequency through the
Goldstone--Ward condition. Within Gaussian theory, the rigid parity
splitting transfers this zero-frequency structure to the copy-odd phase
channel at the finite frequency $2t_h$. The resulting identity
$K_{-P}(2t_h)=0$ is therefore neither a property of the honeycomb
lattice nor an isolated solution of a particular RPA equation. It
follows from the combination of a rigid copy-space splitting and the
Ward identity of the paired state.

This interpretation also clarifies what is and is not universal. For
two identical copies coupled by a momentum-independent scalar $t_h$,
with a common copy-symmetric gap and conventional bandwise singlet
pairing, the pointwise parity-pair identity of \eqn{eqn:pointwise}
survives arbitrary changes of chemical potential, band dispersion,
orbital content, and dimensionality within this class. After momentum
summation, superconducting self-consistency converts that identity into
the Gaussian kernel zero $K_{-P}(2t_h)=0$. The physical collective pole
requires one further condition. At particle-hole symmetry the copy-odd
amplitude decouples, and the kernel zero becomes an isolated
relative-phase pole at $\Omega_L=2t_h$. Away from particle-hole
symmetry the diagonal phase kernel still vanishes at $2t_h$, but the
physical eigenfrequency is determined by the coupled amplitude--phase
sector and is no longer pinned to that value. The distinction between a
kernel zero and a collective pole is therefore intrinsic to the result
rather than a technical qualification.

Pairing does more than generate the Ward anchor. It also isolates the
hybridization scale spectrally. In the $\mathbf q=0$ layer-odd channel
the relevant quasiparticle processes obey
\begin{equation}
  |E_A-E_B|<2t_h<E_A+E_B ,
  \label{eqn:twosided}
\end{equation}
for every $\Delta_0>0$. Thermally activated inter-parity scattering is
confined below $2t_h$, whereas inter-parity pair creation begins above
it. The Gaussian pole at particle-hole symmetry therefore occupies a
genuine two-sided spectral window throughout the paired state. This is
stronger than the usual statement that a collective mode lies below a
pair-breaking threshold. Heating generates quasiparticle scattering on
the low-frequency side, but the rigid hybridization scale remains
separated from both classes of excitations as long as the gap is
finite. Whether the two boundaries actually approach $2t_h$ when
$\Delta_0\to0$ depends on the normal-state band structure. The
half-filled honeycomb spectrum realizes both limiting conditions, so
its spectral window closes onto $2t_h$ at the mean-field transition
while the pole residue simultaneously vanishes. The finite-temperature
pinning is thus a property of the paired Gaussian dynamics and not a
zero-temperature coherence-factor coincidence.

The layer-imbalance response gives this structure a direct microscopic
meaning. The operator $n_-=N_1-N_2$ is the transverse coordinate of the
copy pseudospin and couples to the relative-phase oscillation through
the surviving phase--density susceptibility. At particle-hole symmetry
the Gaussian response reduces to a single pole at $2t_h$. More
strongly, that pole exhausts the Gaussian first moment
$S_1^{\rm G}=-4\langle H_\perp\rangle_{\rm G}$. The full Hamiltonian
obeys the exact operator identity $S_1=-4\langle H_\perp\rangle$,
independently of Gaussian theory. These two statements give a useful
physical separation. Gaussian theory predicts not only a frequency but
also a very specific organization of oscillator strength, while the
exact sum rule constrains the total energy-weighted response even after
interactions redistribute the spectrum. An interlayer bias followed by
a layer-imbalance measurement therefore probes a fixed microscopic
observable rather than an auxiliary pairing-field coordinate.

The many-body interaction changes this picture in a physically
transparent way. The copy pseudospin undergoes a closed Larmor
precession under the kinetic Hamiltonian, but the Hubbard interaction
generates the torque $R=[H_U,L]$, which couples interlayer transfer of
one spin species to the fluctuating layer occupation of the opposite
spin. The local many-body environment therefore enters directly into
the pseudospin dynamics. The uniform hybridization field no longer
produces a closed equation of motion, and there is no exact many-body
reason for the excitation to remain at $2t_h$. The exactly solvable
rung makes this failure explicit. The excitation evolves continuously
from $2t_h$ at $U=0$ toward the pair-transfer scale $4t_h^2/|U|$ at
strong attraction. This crossover has a simple physical
interpretation. Once fermions bind strongly into pairs, relative-layer
dynamics are governed increasingly by virtual pair transfer rather than
by the bare single-particle bonding--antibonding splitting.

The finite-cluster Hubbard spectrum shows that this interaction-driven
detuning survives when intralayer motion is restored. At $|U|=4t$, the
dominant layer-imbalance excitation is already reduced to
$0.467\,(2t_h)$, yet it retains $98.6\%$ of the ordinary spectral
weight. Its share of the exact first moment is smaller, $92.4\%$, and
falls further with increasing attraction. The many-body effect is
therefore not simply the disappearance or broad fragmentation of the
layer-odd excitation. A single line can remain overwhelmingly dominant
in the spectrum while its frequency is strongly renormalized and a
small amount of weight transferred to higher energies carries a
disproportionately large fraction of the energy-weighted sum rule.
Frequency pinning, spectral dominance, and first-moment exhaustion are
distinct physical properties, and interactions need not destroy them
simultaneously.

This distinction also resolves the different senses in which exact
statements enter the problem:
\begin{equation}
\begin{array}{rcl}
  \text{rigid }2t_h\text{ splitting}
    &:& \text{exact one-body identity},\\[2pt]
  K_{-P}(2t_h)=0
    &:& \text{exact within pairing-field}\\
    & & \text{Gaussian theory},\\[2pt]
  S_1=-4\langle H_\perp\rangle
    &:& \text{exact for the full}\\
    & & \text{interacting Hamiltonian}.
\end{array}
  \label{eqn:hierarchy}
\end{equation}
The interacting collective frequency is none of these. At particle-hole
symmetry the first two ingredients combine to produce the Gaussian pole
at $2t_h$, but the full interaction is free to renormalize that pole
while remaining constrained by the third identity. The resulting
physics is therefore a hierarchy rather than a single protected number.
Hybridization supplies a microscopic copy-space scale, superconducting
self-consistency promotes it to an exact Gaussian phase-kernel zero,
pairing isolates that scale from quasiparticle excitations, and the
many-body interaction converts the bare precession into correlated
dynamics without necessarily destroying the dominant layer-odd
excitation.

The present analysis also identifies the boundaries of this mechanism.
Momentum-dependent hybridization removes the single rigid parity
splitting. Inequivalent copies, unequal gaps, copy-antisymmetric
one-body fields, or pairing structures that are not bandwise
conventional singlets remove the algebra underlying
\eqn{eqn:pointwise}. Finite momentum likewise connects different
crystal momenta and therefore does not retain the pointwise
$\mathbf q=0$ cancellation. These cases should generally produce
model-dependent relative-phase dynamics rather than a universal $2t_h$
Gaussian scale. Charged bilayers introduce an additional complication
because long-range Coulomb interactions dress the antisymmetric density
sector. The present work concerns the neutral short-range problem,
where the role of hybridization can be isolated without that
electrodynamic dressing.

The principal unresolved question is consequently not whether the bare
$2t_h$ pinning is an exact many-body theorem. The torque, interacting
rung, and finite-cluster spectrum already show that it is not. What
remains is the interacting relative-phase dynamics in the thermodynamic
symmetry-broken state. Determining its renormalized frequency, spectral
weight, and linewidth requires a treatment beyond Gaussian order that
preserves the collective phase dynamics while incorporating the
interaction-generated torque. Such a calculation would connect the
exact Gaussian reference established here to the full many-body
collective mode and determine how the crossover from
hybridization-controlled precession to correlated pair transfer
develops in an extended system.

\acknowledgments
Y.P.\ thanks Vijay B.\ Shenoy, Manjari Gupta, and Syed R.\ Hassan for
many fruitful discussions. Y.P.\ acknowledges financial support from
the CSIR, India. This work is also supported by the National Research
Foundation of Korea
(NRF), funded by the Ministry of Science and ICT (MSIT), Korea, under
Grants No.\ NRF-2022R1A2C1011646, NRF-2022M3H3A1085772,
RS-2024-00416036, RS-2025-03392969, and RS-2026-25490114, by the
Creation of the Quantum Information Science R\&D Ecosystem program
through the NRF funded by the Korean government (MSIT) (Grant No.\
RS-2023-NR068116), and by the Quantum Simulator Development Project
for Materials Innovation through the NRF funded by MSIT, Korea (Grant
No.\ RS-2023-NR119931).

\section*{Data availability}
The numerical data shown in the figures, and the scripts
that regenerate them, are available from the author upon
reasonable request.

\bibliographystyle{apsrev4-2}
\bibliography{Ref,Ref_SM_v01,Ref_lit_v01}

\clearpage
\onecolumngrid

\setcounter{section}{0}
\setcounter{equation}{0}
\setcounter{figure}{0}
\setcounter{table}{0}
\renewcommand{\thesection}{S\arabic{section}}
\renewcommand{\theequation}{S\arabic{equation}}
\renewcommand{\thefigure}{S\arabic{figure}}
\renewcommand{\thetable}{S\arabic{table}}

\begin{center}
{\Large\bfseries Supplemental Material for\\[2pt]
``Exact Gaussian pinning of a relative-phase mode and\\
interaction-driven detuning from the pinned frequency''}\\[1ex]
Yogeshwar Prasad
\end{center}

This Supplemental Material collects thirteen items referenced from the main
text: the BdG/Nambu convention and the failure mode of the conjugated
hole block (\sect{sm:nambu}); the $t_h=0$ monolayer benchmark
(\sect{sm:monolayer}); the pseudospin structure and coupled-channel
forensics, including the two exact Ward anchors that forbid an
independent resummation of the layer-odd density (\sect{sm:sectors});
the support of the layer-odd bubbles, the frequency scan and the
independent numerical cross-checks (\sect{sm:checks}); the doping and
finite-momentum dependence of the coupled pole (\sect{sm:doping}); the
normalization, sum-rule excess and coupling dependence of the
layer-imbalance residue (\sect{sm:observable}); the status of the
interaction torque and the exact interacting rung benchmark
(\sect{sm:torque}); the explicit Bloch Hamiltonian
(\sect{sm:bloch}); the collective-channel counting and the
normalization convention (\sect{sm:channels}); the derivation of the
pointwise identity (\sect{sm:derivation}); the conjugate lemma for the
coherence factors (\sect{sm:amplitude}); the direct derivation of
sum-rule exhaustion (\sect{sm:exhaustion}); and the exact
diagonalization of the full Hubbard Hamiltonian on a twelve-site
cluster (\sect{sm:ed}). Equations, figures and
sections numbered with the prefix S refer to this document; all others
refer to the main text. Throughout, $t=1$ sets the energy unit, the
reference point is $|U|=4t$, $t_h=0.6t$ at half filling ($\mu=0$), and
the self-consistent gap there is $\Delta_0=1.3545\,t$.

\section{BdG convention: the Nambu hole block must be $-H_0(\mathbf k)$,
not $-H_0(\mathbf k)^*$}
\label{sm:nambu}

A useful implementation check concerns the Nambu hole block, because a
wrong choice silently corrupts the numerical value of the bubble
identity while leaving the identity itself intact. On-site singlet pairing couples $(\mathbf k,\uparrow)$ with
$(-\mathbf k,\downarrow)$, so with the Nambu spinor
$(c_{\mathbf k\uparrow},c^\dagger_{-\mathbf k\downarrow})$ the hole
block of the BdG matrix is
$-H_0(-\mathbf k)^{\mathsf T}=-H_0(\mathbf k)$, using
$f(-\mathbf k)=f(\mathbf k)^*$ and Hermiticity. The superficially
similar choice $-H_0(\mathbf k)^*$ describes a different problem: it
produces spurious subgap quasiparticle states (minimum excitation
$0.24\,t$ instead of $\Delta_0=1.35\,t$ at $|U|=4t$) and shifts the
common value of the two bubbles from the Ward value $-8/|U|$ to an incorrect $-1.783$, an $11\%$ error. The failure is silent: the
wrong hole block still satisfies the pointwise identity
$\chi^{\rm raw}_{-P}(2t_h,\mathbf k)=\chi^{\rm raw}_{+P}(0,\mathbf k)$
exactly. That is not because a wrong BdG problem inherits a physical
operator theorem; the reason is structural. The erroneous implementation
still carries the rigid parity splitting $\xi_A-\xi_B=2t_h$ on which the
algebraic cancellation rests, and that splitting is the only spectral
input the cancellation needs. The same corrupted energies then enter
both sides, so the equality survives while their common value moves off
the Ward value $-8/|U|$.

Equality of the two bubbles is therefore \emph{not} by itself evidence
that an implementation is correct; the Ward-identity value $-8/|U|$ and
the analytic quasiparticle spectrum are the checks that discriminate. The numerical core used for all figures of
the main text asserts $\min_{\mathbf k}|E(\mathbf k)|=\Delta_0$ as a
guard against exactly this substitution.

\section{Monolayer limit and comparison with prior work}
\label{sm:monolayer}

Setting $t_h=0$ decouples the bilayer into two independent honeycomb
monolayers. Because the single-layer honeycomb density of states
vanishes linearly at the Dirac points, the monolayer attractive Hubbard
model has a finite critical coupling $U_c$ for $s$-wave pairing, in
contrast to the bilayer's $U_c=0$. This monolayer limit has been studied
directly: Zhao and Paramekanti~\cite{ZhaoParamekanti2006} analyzed the
BCS--BEC crossover of the attractive Hubbard model on the honeycomb
lattice within mean-field theory, finding a semimetal-to-superfluid
quantum phase transition at a finite mean-field coupling
$U_c^{\rm MF}/t\approx2.2$--$2.3$; the superfluid density of the same
monolayer model was computed in
Ref.~\cite{Iskin2019HoneycombStiffness}. On the quantum Monte Carlo side, Lee
\emph{et al.}~\cite{Bouadim2009} reported an early determinant QMC
estimate $5.0<U_c/t<5.1$ at half filling; that value has since been
superseded. The bipartite half-filled lattice maps the attractive
semimetal-to-(superfluid $+$ density-order) transition exactly onto the
repulsive semimetal--antiferromagnet transition, for which large-scale
QMC gives $U_c/t=3.85\pm0.02$ with Gross--Neveu criticality and
$\nu=1.02(1)$~\cite{Otsuka2016}; Kennedy
\emph{et al.}~\cite{Kennedy2025} adopt that value and state the
attractive-side mirror explicitly. We use $|U_c|/t\approx3.85$ below.
The mirrored phase carries simultaneous $s$-wave pairing and density
order, as required by the pseudospin symmetry discussed in the main
text.

We have checked the mean-field number directly: solving our own gap
equation at $t_h=0$, $T\to0$ (independently of, and with a different
Brillouin-zone mesh than, the bilayer calculations, $N_k=6.4\times10^5$
points) gives $\Delta_0(U)=0$ for $U/t\le2.20$ and $\Delta_0(U)>0$ for
$U/t\ge2.24$; bisection on this threshold gives $U_c^{\rm MF}/t=2.232$,
in agreement with Ref.~\cite{ZhaoParamekanti2006}. This is a nontrivial
consistency check, since the monolayer limit exercises only the diagonal
($t_h=0$) part of the same band structure and gap equation used
throughout the main text, with no parameters retuned. The two quantum Monte Carlo determinations quoted above differ
substantially from each other, and we use the more recent value
$|U_c|/t\approx3.85$ below. The factor $\sim\!1.7$ between the
mean-field threshold and that critical coupling quantifies how strongly quantum
fluctuations renormalize the semimetallic pairing threshold, and should
be kept in mind as the expected level of quantitative (not structural)
mean-field error in this regime. Correlations in the bilayer band
insulator have been studied at finite temperature in
Ref.~\cite{Prasad2022}, and the interplay of disorder and interactions
in the same geometry by determinant quantum Monte Carlo in
Ref.~\cite{Prasad2024}.

\section{Coupled-channel forensics: which sectors mix, and why the
layer-odd density may not be resummed independently}
\label{sm:sectors}

\subsection{Pseudospin SU(2) and the SC/CDW degeneracy}

The degeneracy is shown directly in Fig.~3(a) of the main text: the pairing and bilayer-staggered charge-order
susceptibilities lie on top of each other over the whole temperature
range and satisfy the Thouless condition simultaneously, for every
interlayer hopping shown. The two orders are therefore degenerate
members of one multiplet at half filling, and the paired saddle used
throughout is one representative of it. The degeneracy fixes which
channels can mix, but by itself says nothing about the layer-odd
frequency, which is the subject of \sect{sm:derivation}.At half filling the bipartite structure of the honeycomb lattice endows
the Hamiltonian with an exact pseudospin SU(2)
symmetry~\cite{Yang1989,Zhang1990}. On the four-sublattice partition
$\mathcal A=\{A_1,B_2\}$, $\mathcal B=\{B_1,A_2\}$ with sign
$\epsilon_i=+1$ on $\mathcal A$ and $\epsilon_i=-1$ on $\mathcal B$, the
particle-hole (Shiba) transformation~\cite{Shiba1972} $c_{i\up}\to c_{i\up}$,
$c_{i\dn}\to\epsilon_i c^\dagger_{i\dn}$ leaves the kinetic term
invariant and maps the attractive interaction to its repulsive
counterpart. Under the same map the on-site pairing operator rotates
into the staggered density $\epsilon_i n_i$, so the uniform $s$-wave
paired state and a bilayer-staggered charge-density-wave state --- the
pattern $M_{\rm CDW}={\rm diag}(+1,-1,-1,+1)$ on the
$(A_1,B_1,A_2,B_2)$ basis --- are exact pseudospin partners at half
filling: $\chi_{\rm pair}(T)=\chi_{\rm CDW}(T)$ identically for every
$t_h$, and the two channels' Thouless temperatures coincide. We
verified this numerically for $t_h\le0.6\,t$: at $t_h=0,\,0.3,\,0.6\,t$
the two susceptibilities agree to round-off
(relative) at $T=0.3t$. We have not tested larger splittings on a mesh
fine enough to separate a genuine violation from finite-mesh
discretization, and make no numerical claim there; the identity itself
is a symmetry statement and holds for every $t_h$.

\subsection{Which cross-bubbles vanish}

At the particle-hole-symmetric saddle the layer-odd phase vertex has
\emph{exactly} vanishing cross-bubbles (numerically zero to round-off) with the
layer-symmetric density, with the staggered-CDW density pattern
$(+,-,-,+)$, and with both amplitude channels. The pseudospin-partner
CDW sector therefore does not mix with the Leggett channel at the
Gaussian level, and the SU(2) structure of the main text carries no
hidden fluctuation coupling into the $2t_h$ result. We emphasize that
this statement concerns the CDW \emph{pattern} only: the layer-odd
density does not decouple. The single nonvanishing cross-bubble is with
the layer-odd density, $\chi_{\phi_-n_-}(2t_h)$, which is purely
imaginary and of order unity. This is not a defect but the Larmor
structure made visible --- the physical layer-odd oscillation is a
canonically conjugate relative-phase/relative-density precession, the
fluctuation image of the operator $L=T_z-iT_y$ --- and it is precisely
why a layer-imbalance probe couples to the mode, as quantified by the
sum rule in the main text.

\subsection{Two exact Ward anchors forbid an independent density
resummation}

One might worry that an order-unity phase--density coupling shifts the
collective pole off $2t_h$ once the density channel is treated as an
independent fluctuation field. Two exact Ward limits show that the naive
version of that extension is not admissible.

First, at half filling the staggered-CDW kernel possesses its own exact
zero, $\chi_{\rm stag}(0)=-8/|U|$ (verified to machine precision), the
Goldstone partner demanded by pseudospin SU(2) --- and the staggered
density has an equally large quadrature cross-bubble to the staggered
pair phase, the exact SU(2) image of the layer-odd coupling. Second, at
$t_h\to0$ the layer-odd phase becomes an independent Goldstone mode
(each layer carries its own U(1)), yet the layer-odd phase--density
cross-bubble remains of order unity in that limit. In both cases an
exact symmetry pins the pole at $\omega=0$.

A naive resummation that treats the layer-odd density as a second,
independently Hubbard--Stratonovich-resummed Gaussian field with its own
interaction vertex shifts \emph{both} of these symmetry-pinned Goldstone
poles --- we verified this explicitly --- and is therefore an
inconsistent mixed-channel (Fierz-type) construction. The simultaneous
failure of the two exact limits is the decisive evidence, with the
double-counting structure explaining \emph{why} the failure occurs.
Consequently its shifted pole cannot be used to modify the pairing-field
Gaussian theorem, which is established independently by the pointwise
identity. The theorem of the main text is a statement about that
consistently-defined single-channel Gaussian theory, in which the
density coupling determines the \emph{character} of the mode --- its
precession plane and its observability --- not its Gaussian frequency.
The physical layer-imbalance response built on this same single-channel
theory, and the exact f-sum rule it satisfies, are given in the main
text.

\section{Frequency scan and independent numerical cross-checks}
\label{sm:checks}

Unless stated otherwise, the spectral functions shown in the main text are evaluated
on an $N_{\rm BZ}=90$ Brillouin-zone mesh with Lorentzian broadening $\eta=0.02\,t$;
the mean-field gap and $T_c$ quoted there use a $500\times500$ mesh.

\subsection{Method}

The identity was independently verified with a direct $8\times8$ BdG
Lehmann calculation:
building the $8\times8$ BdG matrix at each of $N_k=250{,}000$
Brillouin-zone points at the self-consistent $\Delta_0=1.3545\,t$
($|U|=4t$, $t_h=0.6t$), diagonalizing exactly, and summing the retarded
Lehmann bubble for the symmetric-phase vertex at $\omega\to0$ and the
antisymmetric-phase vertex at $\omega=2t_h$ directly from the resulting
eigenvectors and eigenvalues, with no closed-form bubble formula
assumed. The figures of the main text use the same construction on
coarser meshes.

\subsection{The coincidence is frequency-specific}

A frequency scan confirms that the equality of the two bubbles is a
genuine, isolated coincidence rather than a trivial identity holding at
all $\omega$: the two sides differ by $0.063$ at $\omega=1.0t$, by
$0.034$ at $\omega=1.1t$, coincide exactly at $\omega=1.2t=2t_h$, and
diverge again to $0.038$ at $\omega=1.3t$.

\subsection{Support of the layer-odd bubbles}
\label{sm:support}

The layer-odd vertex is $\lambda_z$ in copy space and the identity in
the orbital (sublattice) space. In the bonding/antibonding basis
$\lambda_z$ is purely off-diagonal, so the vertex is parity-odd; being
the identity on the orbital index it is simultaneously band-diagonal.
Its two-quasiparticle transitions therefore connect parity partners of
the \emph{same} band, whose energies differ by exactly $2t_h$ in
$\xi$ --- the same kinematic input as the theorem --- and the support
of every bubble built from the sublattice-uniform layer-odd vertex
$\lambda_zs_0$ begins at the inter-parity threshold $\omega_c^-$, not
at the intra-parity edge $2\Delta_0$. The restriction to that vertex is
necessary: the sublattice-staggered layer-odd channel $P{-}v$ has
weight starting at $2\Delta_0$ (\sect{sm:channels}).

We confirmed this directly by inspecting the Lehmann transition list at
$|U|=4t$, $t_h=0.6t$ on a $90\times90$ mesh. For the layer-odd density,
the layer-odd phase, and their cross bubble, the smallest transition
energy carrying nonzero weight is $2.9636\,t=\omega_c^-$ in every
case, and the total weight in the window
$[2\Delta_0,\omega_c^-)=[2.709,2.964)\,t$ is \emph{exactly zero} (no
transitions, to round-off). The layer-\emph{even}
phase bubble, by contrast, has its support starting at
$2\Delta_0=2.709\,t$ as expected for an intra-parity channel. Any
apparent weight below $\omega_c^-$ in a plotted sublattice-uniform
layer-odd spectral function is Lorentzian broadening of the collective
pole, not spectral weight.

\subsection{Three cross-checks}

(i) Recomputing $\chi^{\rm raw}_{-P}(2t_h)$ after rotating to the
bonding/antibonding (layer-parity) basis, in which the antisymmetric
vertex connects the two parity blocks, agrees with the direct
layer-basis calculation to round-off.

(ii) Amplitude--phase orthogonality in the antisymmetric channel,
$\chi_-^{(\rm amp,phase)}(\omega)\equiv0$, required by particle-hole
symmetry at half filling, holds to round-off (\sect{sm:doping}).

(iii) The quasiparticle spectrum returned by the numerical
diagonalization coincides with the analytic BdG dispersion
$E_{\sigma}(\mathbf k)=[(|f(\mathbf k)|+\sigma
t_h)^2+\Delta_0^2]^{1/2}$ to machine precision, with minimum
$\Delta_0$. At the reference point this places the intra-parity
(layer-even) two-quasiparticle onset at $2\Delta_0=2.709\,t$ and the
inter-parity (layer-odd) onset --- the one the Leggett vertex actually
couples to --- at $\omega_c^-=2\sqrt{t_h^2+\Delta_0^2}=2.963\,t$, both
far above $2t_h=1.2\,t$.

\subsection{Independent verification of the pointwise identity}
\label{sec:verify}

\begin{figure*}[t]
  \centerline{\includegraphics[width=\textwidth]{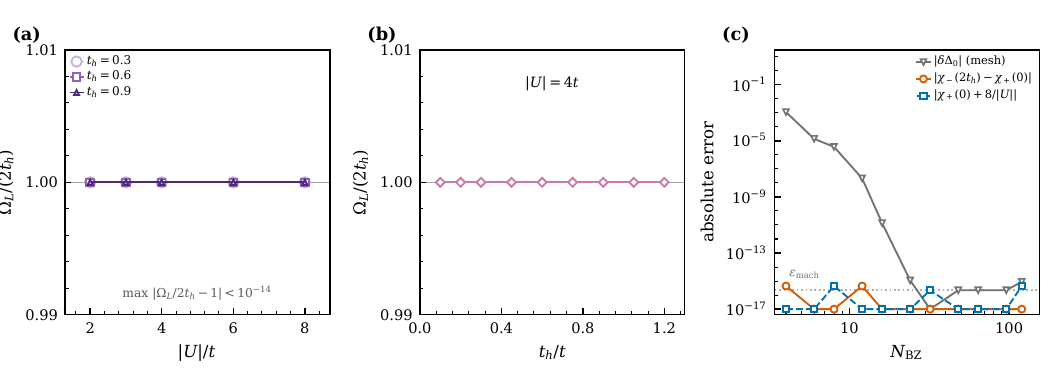}}
  \caption{Exactness of the layer-odd kernel zero, from the
  $8\times8$ BdG Lehmann construction with $\Delta_0$ solved
  self-consistently on the same mesh used for the bubbles.
  (a) $\Omega_L/(2t_h)$ versus $|U|/t$ at $t_h/t=0.3,0.6,0.9$ (marker
  sizes nested; the curves coincide). (b) The same ratio versus $t_h/t$
  at $|U|=4t$. Across both panels the ratio is unity to round-off, in
  windows spanning $\pm1\%$; (c) Mesh dependence at $|U|=4t$, $t_h=0.6t$: the identity and Ward
  errors (circles, squares) stay at round-off from $N_{\rm BZ}=4$ to
  $120$, while the gap (triangles) converges only exponentially --- the
  identity holds pointwise, before any quadrature has converged.}
  \label{fig:exactness}
\end{figure*}

An independent $8\times8$ Bogoliubov--de~Gennes Lehmann calculation at
$N_k=250{,}000$ points assuming no closed-form bubble. We obtain
\begin{equation}
  \chi^{\rm raw}_{+P}(0) = \chi^{\rm raw}_{-P}(2t_h) = -\frac{8}{|U|},
  \label{eqn:numresult}
\end{equation}
exactly, and over a grid in $|U|$ and $t_h$ the kernel zero sits at
$\Omega_L/(2t_h)=1$ to round-off with a finite residue
[\fig{fig:exactness}(a,b)].

\fig{fig:exactness}(c) makes the pointwise character visible: on a
$4\times4$ mesh --- sixteen momenta, where the gap itself is still
wrong at $10^{-3}$ --- the identity already holds to round-off. That is
not mesh convergence but
the pointwise identity acting before any quadrature; the Goldstone anchor
is itself discrete, since solving $\Delta_0$ on the same mesh makes the
Ward relation exact there. A Nambu-convention pitfall that corrupts \eqn{eqn:numresult} by $11\%$ while leaving the pointwise identity intact is discussed in \sect{sm:nambu}.

\section{Doping and finite-momentum dependence of the coupled pole}
\label{sm:doping}

The frequency of the full layer-odd mode is the root of the coupled
determinant
\begin{equation}
  \det\begin{pmatrix} K^-_{AA} & K^-_{AP}\\
                      K^-_{PA} & K^-_{PP}\end{pmatrix}=0 ,
  \label{eqn:sm-coupleddet}
\end{equation}
so the diagonal condition $K^-_{PP}(2t_h)=0$ alone does not fix the root
once the antisymmetric amplitude--phase element $K^-_{AP}$ is nonzero.
Particle-hole symmetry forces $K^-_{AP}=0$ (we find
$\chi^-_{(\rm amp,phase)}$ zero to round-off), and the root then sits
exactly at $2t_h$. The condition is the symmetry, not the filling
value: for the bipartite honeycomb the two coincide at $\mu=0$, but a
band without particle-hole symmetry has $K^-_{AP}\neq0$ already at
$\mu=0$ (\sect{sm:exhaustion}).

Numerically, $K^-_{AP}$ vanishes at the particle-hole-symmetric point to
round-off and switches on \emph{linearly} for sufficiently small
detuning. Because the off-diagonal terms enter
\eqn{eqn:sm-coupleddet} only through the product $K^-_{AP}K^-_{PA}$,
the displacement of the root is quadratic,
$\Omega_{\rm full}-2t_h=O(\delta n^2)$, which is the statement made in
the main text. We retain only this local structural result
$K^-_{AP}=O(\delta n)$; no finite doping interval or numerical
coefficient is inferred from an asymptotic statement of this kind.

\subsection{The finite-momentum curvature is not extracted}

Analyticity of the kernel about $(\omega,\mathbf q)=(2t_h,\mathbf 0)$
together with $C_{6v}$ symmetry constrains the dispersion to
$\Omega_L^2(\mathbf q)=(2t_h)^2+c_L^2|\mathbf q|^2+O(q^4)$, isotropic at
quadratic order. Finite-difference estimates of $c_L^2$ along
$\Gamma\to M$ and $\Gamma\to K$ differ by approximately $12\%$, although
$C_{6v}$ symmetry requires equality at this order. The available
momentum grid is therefore insufficient for a quantitative extraction of
$c_L^2$, and we quote no numerical curvature. The symmetry statement
stands on its own; nothing elsewhere in this work depends on the value
of $c_L$.

\section{Layer-imbalance response: normalization, bare-bubble excess,
and the coupling dependence of $R_-$}
\label{sm:observable}

Everything in this section is evaluated on the self-consistent Gaussian
saddle. We write $\langle\cdot\rangle_{\rm G}$ for the saddle
expectation and
$S_1^{\rm G}\equiv-4\langle H_\perp\rangle_{\rm G}$ for the first moment
of the Gaussian response, reserving $S_1=-4\langle H_\perp\rangle$ for
the exact interacting moment used in \sect{sm:ed}. The operator
identity behind both is the same; only the state in which
$\langle H_\perp\rangle$ is evaluated differs, and the two coincide only
at $U=0$.

\emph{Normalization.} $n_-=N_1-N_2$ is the extensive layer-imbalance
operator and $\chi^R_{--}$ is taken per unit cell (four sites, two per
layer), so $R_-$ has units of energy. The dimensionless ratio
$R_-/S_1^{\rm G}$
is invariant under $n_-\to a\,n_-$ and is the quantity that carries
physical meaning; $R_-$ alone does not, unless the normalization is
quoted with it, and the kernel residue $Z_L$ does not at all.

\emph{The bare bubble violates the sum rule; the vertex correction
repairs it.} Evaluated on the mean-field saddle, the bare BdG bubble
gives a first moment $-4\langle H_\perp\rangle_{\rm G}+32\Delta_0^2/|U|$
(verified to machine precision). The excess is the
particle-number-violating piece
$\langle[n_-,[H_\Delta,n_-]]\rangle$: the mean-field pairing term does
not commute with $n_-$, whereas the full Hamiltonian's interaction
does. The RPA vertex correction removes it exactly, which is the
standard statement that a conserving Gaussian response obeys the sum
rules of the full Hamiltonian rather than those of the mean-field
Hamiltonian. Numerically, $\omega^2\chi_{--}(\omega)$ tends to
$1.56997$ as $\omega\to\infty$, matching
$-4\langle H_\perp\rangle_{\rm G}$ to round-off, while the bare bubble tends to
$16.2476$.

\emph{The zero is a simple pole.} The kernel residue
$|Z_L|=|\partial K_{-P}/\partial\omega^2|^{-1}$, evaluated at
$\omega=2t_h$, is finite at every coupling tested, so the zero of the
layer-odd phase kernel is simple and the response has a pole rather
than a higher-order singularity there. Its \emph{value} is a
Hubbard--Stratonovich convention --- it rescales with the three kernel
normalizations related above --- and carries no physical meaning on its
own; the physical weight is $R_-$, and the convention-free number is
$R_-/S_1^{\rm G}$.

\emph{Single-pole tests.} Two independent checks confirm that the
dressed response is a single pole and not merely a pole that happens to
saturate the first moment. (i) The dressed spectral weight above
$\omega_c^-$ scales linearly with the broadening $\eta$ --- at
$\omega=3.3\,t$ it falls linearly from $7.4\times10^{-4}$ to
$1.8\times10^{-6}$ as $\eta/t$ is reduced from $2\times10^{-2}$ to
$5\times10^{-5}$ --- so it is
Lorentzian leakage from the pole and vanishes as $\eta\to0$, whereas
the bare bubble retains a finite $7.7\times10^{-2}$ there. (ii) The
third moment satisfies $S_3^{\rm G}/S_1^{\rm G}\to(2t_h)^2=1.4400$ to seven digits, the
signature of a single pole, at every coupling tested.

\emph{Coupling dependence.} Because $R_-=-4\langle H_\perp\rangle_{\rm G}$
within the Gaussian theory, the absolute residue simply tracks the
interlayer kinetic energy of the saddle. It is nonmonotonic in the coupling,
rising from $0.921\,t$ at $|U|=2t$ to $1.570\,t$ at $|U|=4t$ and
falling to $1.211\,t$ at $|U|=8t$ (at $t_h=0.6t$), and it vanishes as
$t_h\to0$, where $n_-$ becomes a conserved quantity and the mode must
carry no weight at all. The ratio $R_-/S_1^{\rm G}$ remains unity to round-off
across the whole range.

\section{Status of the interaction torque, and the exact interacting
rung benchmark}
\label{sm:torque}

\subsection{The torque operator}

The torque $R=[H_U,L]$ of the main text is exact and nonvanishing, and
it is irreducibly two-body: it has zero Hilbert--Schmidt overlap with
every layer-pseudospin generator and with $L$ itself. That is the whole
of what we establish about it. The closed form quoted in the main text
is an analytic operator identity; it was additionally verified against
the commutator evaluated directly in the full Fock space of a four-site
cluster, which reproduces it term by term.

The surviving one-body Wick contractions show that the operator
structure of $R$ alone fixes neither the perturbative order of the
frequency correction nor a decay threshold. In particular, contractions
proportional to
$\langle c^\dagger_{j1\sigma}c_{j2\sigma}\rangle\propto\langle
T_x\rangle\neq0$ generate layer-odd density ($n_-$) bilinears, while
anomalous contractions generate interlayer pair bilinears. The Gaussian
theorem of \sect{sm:derivation} and the sum-rule exhaustion of
\sect{sm:exhaustion} are independent of $R$.

Quantitative frequency shifts and linewidths therefore require a
beyond-Gaussian treatment of the fluctuation effective action, through
the one-loop self-energy of the collective mode constructed from its
cubic and quartic vertices about the saddle.

\subsection{Exact interacting rung benchmark}
\label{sm:dimer}

The torque establishes that nothing protects $2t_h$ beyond Gaussian
order, but not what replaces it. One case can be solved exactly at every
coupling: a single vertical rung, i.e.\ two sites coupled only by $t_h$,
with the same on-site attraction.

In the two-particle spin-singlet sector, write
$|D_\pm\rangle=(|{\up\dn},0\rangle\pm|0,{\up\dn}\rangle)/\sqrt2$ for the
symmetric and antisymmetric doubly occupied states and
$|S\rangle=(|{\up},{\dn}\rangle-|{\dn},{\up}\rangle)/\sqrt2$ for the
singly occupied singlet. In the basis $(D_+,S,D_-)$, and up to a
constant,
\begin{equation}
  H_{\rm rung}=
  \begin{pmatrix}
    -|U| & -2t_h & 0\\
    -2t_h & 0 & 0\\
    0 & 0 & -|U|
  \end{pmatrix} ,
  \label{eqn:sm-rungH}
\end{equation}
so $E_{\rm GS}=-\tfrac12\big(|U|+\sqrt{U^2+16t_h^2}\big)$ while
$E_{D_-}=-|U|$. The layer-imbalance operator $T_z$ interchanges $D_+$
and $D_-$, so $D_-$ is precisely the layer-odd bright state, and the
exact excitation energy is
\begin{equation}
  \Omega_{\rm rung}(U)=E_{D_-}-E_{\rm GS}
  =\frac{\sqrt{U^2+16t_h^2}-|U|}{2} .
  \label{eqn:sm-rung}
\end{equation}
It equals $2t_h$ at $U=0$, decreases monotonically with $|U|$, and
approaches the pair-transfer scale $4t_h^2/|U|$ at strong coupling.
Interactions therefore detune the bare hybridization frequency
immediately and in a definite direction; the Gaussian pinning is not an
all-orders statement, and here the departure is exact rather than
bounded.

We use \eqn{eqn:sm-rung} as an exact full-interaction benchmark, not as a
prediction for the thermodynamic lattice mode. On a lattice the
layer-odd response also involves intralayer pair dynamics. Section
\ref{sm:ed} restores that dynamics exactly on a finite periodic cluster;
what remains outside the scope of this work is the thermodynamic-limit
collective mode, for which no extrapolation is attempted here.

\section{Explicit Bloch Hamiltonian}
\label{sm:bloch}

In the four-sublattice Bloch basis $(A_1,B_1,A_2,B_2)$ the normal-state
Hamiltonian of the main text is
\begin{equation}
  h(\mathbf k) = \begin{pmatrix}
    0 & t f(\mathbf k) & -t_h & 0 \\
    t f^*(\mathbf k) & 0 & 0 & -t_h \\
    -t_h & 0 & 0 & t f(\mathbf k) \\
    0 & -t_h & t f^*(\mathbf k) & 0
  \end{pmatrix} - \mu\,\openone ,
  \label{eqn:sm-hk}
\end{equation}
with $f(\mathbf k)=-\sum_{\bm\delta}e^{i\mathbf k\cdot\bm\delta}$ the
dimensionless honeycomb structure factor summed over the three
nearest-neighbor bond vectors. Rotating to the bonding/antibonding
basis block-diagonalizes \eqn{eqn:sm-hk} into two standard two-band
honeycomb problems with eigenvalues
$\varepsilon_{\sigma,s}(\mathbf k)=\sigma t_h+s\,t|f(\mathbf k)|-\mu$,
$\sigma,s=\pm1$, which we verified numerically against \eqn{eqn:sm-hk}
to round-off across the Brillouin zone.

\section{Collective-channel counting and the normalization convention}
\label{sm:channels}

Every prefactor in the sections that follow --- the coherence factors of
\sect{sm:derivation}, the positivity bound of \sect{sm:amplitude}, the
residue and the sum rule of \sect{sm:exhaustion} --- rests on how many
independent collective channels the Gaussian theory contains and on how
each of them is normalized. We derive both here, once, so that the later
prefactors can be traced to a single convention instead of being
restated channel by channel. One warning on labels: in the channel names
below $A$ and $P$ stand for amplitude and phase, while the subscripts
$A,B$ on $\xi$ and $E$ continue to denote the two interlayer-parity
partners of \sect{sm:derivation}, as they do everywhere else in this
document.

\subsection{The field space and its dimension}

The on-site attraction decouples in one channel only, with a single
complex pairing field per site of the unit cell. The AA-stacked bilayer
honeycomb cell contains $N_{\rm orb}=4$ sites $(A_1,B_1,A_2,B_2)$, so
the fluctuation $\delta\Delta_{l\alpha}$ about the uniform saddle has
four complex --- that is, \emph{eight real} --- components at every
$(\mathbf q,\omega)$. This is the complete fluctuation field space
generated by this pairing-channel Hubbard--Stratonovich
representation. The count is set by the number of sites in the cell and by
nothing dynamical, so it is the same at every momentum and frequency; a
counting that returns more channels has added a field the
Hubbard--Stratonovich transformation did not produce (the layer-odd
density of \sect{sm:sectors} is exactly such an addition, and
\sect{sm:sectors} shows what goes wrong when it is added), and one that
returns fewer has discarded a component that the kernel in general
mixes.

We resolve the orbital index in the basis of sign patterns
\begin{equation}
\begin{split}
  \epsilon^{+u}&=(+,+,+,+),\qquad \epsilon^{+v}=(+,-,+,-),\\
  \epsilon^{-u}&=(+,+,-,-),\qquad \epsilon^{-v}=(+,-,-,+),
\end{split}
  \label{eqn:sm-patterns}
\end{equation}
listed on $(A_1,B_1,A_2,B_2)$ and equal to $\lambda_0s_0$,
$\lambda_0s_z$, $\lambda_zs_0$ and $\lambda_zs_z$ respectively, with
$\lambda$ acting on the layer index and $s$ on the sublattice index. The
superscript records the layer parity and $u,v$ the sublattice character,
uniform or staggered. The four patterns are orthogonal and
$\epsilon^c/2$ is orthonormal,
$\sum_{l\alpha}(\epsilon^c/2)_{l\alpha}(\epsilon^{c'}/2)_{l\alpha}
=\delta_{cc'}$, so writing
\begin{equation}
  \delta\Delta_{l\alpha}
  =\sum_c\big(\varphi^A_c+i\varphi^P_c\big)\,
   \frac{\epsilon^c_{l\alpha}}{2}
  \label{eqn:sm-fieldexp}
\end{equation}
turns the Hubbard--Stratonovich term into
$\sum_{l\alpha}|\delta\Delta_{l\alpha}|^2/|U|
=\sum_c[(\varphi^A_c)^2+(\varphi^P_c)^2]/|U|$. The eight real
coordinates $\varphi_X$, with
$X\in\{A,P\}\times\{+u,+v,-u,-v\}$, are then orthonormal and each
carries the \emph{same} bare weight $1/|U|$. That is the normalization
convention used here and, through the dictionary below, everywhere else
in this work.

\subsection{Vertices, kernel, and the Ward anchor}

With \eqn{eqn:sm-fieldexp} the fluctuation enters the BdG matrix as
$\delta H=\sum_X\varphi_XV_X$ with
\begin{equation}
  V^A_c=\tau_x\otimes\frac{\epsilon^c}{2},\qquad
  V^P_c=-\tau_y\otimes\frac{\epsilon^c}{2},
  \label{eqn:sm-verts}
\end{equation}
orthonormal in the same sense, $\tfrac12{\rm Tr}(V_XV_Y)=\delta_{XY}$,
the $\tfrac12$ undoing the Nambu doubling. Integrating out the fermions
gives the Gaussian kernel
\begin{equation}
  M_{XY}(\mathbf q,\omega)=\frac{\delta_{XY}}{|U|}
  -\Pi_{XY}(\mathbf q,\omega),
  \label{eqn:sm-Mdef}
\end{equation}
\begin{equation}
  \Pi_{XY}=-\frac{1}{2N_k}\sum_{\mathbf k}\sum_{ab}
  \frac{(f_b-f_a)\;V^{X}_{ab}V^{Y}_{ba}}{\omega+E_a-E_b+i0^+},
  \label{eqn:sm-Pidef}
\end{equation}
where $V^X_{ab}=\langle a_{\mathbf k}|V_X|b_{\mathbf k+\mathbf q}\rangle$,
the labels $a,b$ run over the eight BdG eigenstates at $\mathbf k$ and
$\mathbf k+\mathbf q$, and the same $\tfrac12$ appears. In this
normalization $\Pi_{P+u}(\mathbf 0,0)=1/|U|$, so
$M_{P+u}(\mathbf 0,0)=0$: the Goldstone condition and the gap equation
are one statement, and no coupling constant is fitted anywhere. This is
the Ward anchor quoted in \sect{sm:derivation} as
$\chi^{\rm raw}_{+P}(0)=-8/|U|$; the $8$ is not a choice but the product
of the pattern normalization $\tfrac14$ in \eqn{eqn:sm-verts} and the
Nambu $\tfrac12$ in \eqn{eqn:sm-Pidef}.

The three kernel normalizations used below are related by the following
dictionary. Writing
$\chi^{\rm raw}_c$ for the Lehmann bubble taken with the
\emph{unnormalized} vertex $\tau_{x,y}\otimes\epsilon^c$ and without the
Nambu $\tfrac12$ --- the form used in \sect{sm:derivation} and
\sect{sm:exhaustion} --- one has $\Pi_c=-\chi^{\rm raw}_c/8$ and
\begin{equation}
  \underbrace{1-g_0\chi^{\rm raw}_c}_{\text{main text, }\sect{sm:exhaustion}}
  =|U|\,M_c ,\qquad
  \underbrace{\frac{2}{|U|}-\Pi^{\rm SM}_c}_{\sect{sm:amplitude}}
  =2\,M_c ,
  \label{eqn:sm-dict}
\end{equation}
with $g_0=1/\chi^{\rm raw}_{+P}(0)=-|U|/8$ and
$\Pi^{\rm SM}_c=2\Pi_c$. All three are positive multiples of each other,
so they share their zeros, their signs and their pole positions; only
their absolute values differ, and a number quoted from one of them must
be read with its own prefactor. The dimensionless ratio $R_-/S_1^{\rm G}$ of
\sect{sm:observable} is invariant under all three, which is why it, and
not $R_-$ alone, is the quantity we report.

The absolute scale can be fixed without evaluating any bubble. In the
orthonormal basis the saddle is itself a vector: $\Delta_{l\alpha}
=\Delta_0$ on all four orbitals means
$\varphi^A_{+u}=\sqrt{N_{\rm orb}}\,\Delta_0=2\Delta_0$ with the other
seven coordinates zero, and the corresponding BdG operator has
Hilbert--Schmidt norm
\begin{equation}
  \bar\Delta\equiv\big\|\Delta_0\,\tau_x\otimes\openone\big\|_{\rm HS}
  =\sqrt{2N_fN_{\rm orb}}\;\Delta_0=2\sqrt{2N_f}\,\Delta_0 ,
  \label{eqn:sm-Dbar}
\end{equation}
$N_f=1$ being the number of fermion flavors and the remaining $2$ the
Nambu doubling. The two differ only by the universal factor
$\sqrt{2N_f}$, common to all eight channels and therefore absent from
every kernel zero. The useful point is that $\bar\Delta$ counts
channels: the factor $\sqrt{N_{\rm orb}}=2$ appears because the uniform
pattern gathers exactly four orbital components, so a measurement of
$\bar\Delta$ tests \eqn{eqn:sm-patterns} and not merely the value of
$\Delta_0$. The same norm fixes the amplitude of the exact $U(1)$
rotation $\delta\Delta=i\theta\Delta_0$, which moves the field along
$\varphi^P_{+u}$ and along no other coordinate.

\subsection{Selection rules and the block structure}

Two selection rules are exact and a third, often assumed, is not.

\emph{Layer parity.} The interlayer term of \eqn{eqn:sm-hk} is
$-t_h\lambda_x$, independent of $\mathbf k$, so $\lambda_x$ commutes
with $h(\mathbf k)$ at every momentum and bonding/antibonding parity
labels every BdG eigenstate at every $\mathbf q$. Vertices carrying
$\lambda_0$ are parity-diagonal and those carrying $\lambda_z$
parity-off-diagonal, so the product of matrix elements in
\eqn{eqn:sm-Pidef} vanishes term by term whenever $X$ and $Y$ have
opposite layer parity. The rule holds at all $\mathbf q$, all $\omega$
and any filling.

\emph{Amplitude versus phase.} At half filling particle-hole symmetry
forbids $\tau_x$--$\tau_y$ mixing, as already used for $K^-_{AP}$ in
\sect{sm:doping}; away from half filling the mixing switches on linearly
in $\delta n$ and the blocks below fuse in pairs.

\emph{Sublattice character.} The uniform pattern $s_0$ is diagonal in
the sublattice band index while the staggered pattern $s_z$ is purely
off-diagonal in it, because the honeycomb band eigenvectors are
$(1,\pm f^*/|f|)/\sqrt2$ in the sublattice index. At $\mathbf q=0$ both ends of
the bubble are built on the \emph{same} eigenbasis, the two properties
are incompatible, and $\Pi_{uv}$ vanishes pointwise in $\mathbf k$. At
$\mathbf q\neq0$ the eigenbasis rotates by
$\theta_{\mathbf k+\mathbf q}-\theta_{\mathbf k}$ between the two ends
and the incompatibility is lost: $u$ and $v$ mix. The sublattice
exchange $s_x$ implements inversion here,
$s_xh(\mathbf k)s_x=h(-\mathbf k)$, and sends $s_z\to-s_z$, so the mixing
element is odd in $\mathbf q$ and vanishes only at $\mathbf q=0$; the
$\tau$ structure makes it purely imaginary.

Hence the block structure. At particle-hole symmetry and generic
$\mathbf q$, the $8\times8$ Hermitian kernel splits into four $2\times2$
blocks labelled by
(amplitude or phase) $\times$ (layer parity),
\begin{equation}
  (A{+}u,A{+}v)\ \ (P{+}u,P{+}v)\ \ (A{-}u,A{-}v)\ \ (P{-}u,P{-}v),
  \label{eqn:sm-blocks}
\end{equation}
each acting in the two-dimensional sublattice-character space; at
$\mathbf q=0$ these blocks become diagonal and the eight channels
decouple completely. Away from particle-hole symmetry the
amplitude--phase mixing noted above switches on and the blocks fuse in
pairs. Every $\mathbf q=0$ result in this Supplemental Material ---
the pointwise identity, the amplitude positivity bound, the residue and
the sum rule --- is therefore a statement about an individual scalar
channel, and no $2\times2$ diagonalization is implied in any of them.

\subsection{Transitions, and which channels are pinned}

\begin{table*}[t]
\caption{Uniform-channel kernel zeros and continuum anchors: the eight
collective channels at $\mathbf q=0$. $X=(\sigma,s)$
labels the four normal states at $\mathbf k$ with
$\xi_X=\sigma t_h+s|f(\mathbf k)|-\mu$ and $E_X=(\xi_X^2+\Delta_0^2)^{1/2}$;
$\pi(X)$ is the partner state selected by the orbital pattern, and
$\Omega_X=E_X+E_{\pi(X)}$ is the transition energy. The coherence
numerator is $\mathcal N^{\rm ph}$ for a phase ($\tau_y$) vertex and
$\mathcal N^{\rm am}$ for an amplitude ($\tau_x$) vertex, in the notation
of \eqn{eqn:sm-three}, and the weight is $\mathcal N/2p$ with
$p=E_XE_{\pi(X)}$. Onsets are $\min_{\mathbf k}\Omega_X$ at half filling,
with $2\Delta_0=2.7090\,t$ and
$\omega_c^-=2(t_h^2+\Delta_0^2)^{1/2}=2.9629\,t$ at the reference
point, the latter closed form being the half-filled honeycomb value
attained at the Dirac points and not a property of the general
two-copy class. The pinned layer-odd amplitude frequency is a separate
object, $\omega^*_{A-}=2(t_h^2+\Delta_0^2)^{1/2}$, exact for the whole
class by \eqn{eqn:sm-pin}; it obeys $\omega^*_{A-}\le\omega_c^-$ always
and equals it here. Pinned zeros are exact
and follow from \eqn{eqn:sm-pin}; unpinned ones are numerical values on a
$200\times200$ mesh, stable to the digits shown between $120\times120$
and $600\times600$. A pinned zero is a zero of the kernel and not
necessarily an isolated pole: the two pinned phase zeros lie strictly
below their own continua and are modes, whereas the two pinned
amplitude zeros sit on their continuum edge here and are threshold
singularities.}
\label{tab:sm-channels}
\begin{ruledtabular}
\begin{tabular}{lccccll}
Channel & pattern & $\xi_{\pi(X)}$ & $\Omega_X$ & $\mathcal N_X$ &
  onset & $\mathbf q=0$ kernel zero \\
\hline
$A{+}u$ & $\lambda_0s_0$ & $\xi_X$ & $2E_X$ & $2\xi_X^2$ &
  $2\Delta_0$ & $2\Delta_0$ (pinned; on its edge, Higgs) \\
$P{+}u$ & $\lambda_0s_0$ & $\xi_X$ & $2E_X$ & $2E_X^2$ &
  $2\Delta_0$ & $0$ (pinned; Anderson--Bogoliubov) \\
$A{+}v$ & $\lambda_0s_z$ & $2\sigma t_h-\xi_X$ & $E_X+E_{\pi(X)}$ &
  $\mathcal N^{\rm am}$ & $\omega_c^-$ & $2.963\,t$ (unpinned; at its edge
  within resolution) \\
$P{+}v$ & $\lambda_0s_z$ & $2\sigma t_h-\xi_X$ & $E_X+E_{\pi(X)}$ &
  $\mathcal N^{\rm ph}$ & $\omega_c^-$ & $2.3707\,t$ (unpinned; subcontinuum,
  collective) \\
$A{-}u$ & $\lambda_zs_0$ & $\xi_X-2\sigma t_h$ & $E_A+E_B$ &
  $\mathcal N^{\rm am}$ & $\omega_c^-$ & $\omega^*_{A-}$ (pinned; $=\omega_c^-$ here, threshold) \\
$P{-}u$ & $\lambda_zs_0$ & $\xi_X-2\sigma t_h$ & $E_A+E_B$ &
  $\mathcal N^{\rm ph}$ & $\omega_c^-$ & $2t_h$ (pinned; isolated pole, relative phase) \\
$A{-}v$ & $\lambda_zs_z$ & $-\xi_X$ & $2E_X$ & $0$ &
  --- & none at any $\omega$ (null channel) \\
$P{-}v$ & $\lambda_zs_z$ & $-\xi_X$ & $2E_X$ & $2\Delta_0^2$ &
  $2\Delta_0$ & $2.0295\,t$ (unpinned; subcontinuum, collective) \\
\end{tabular}
\end{ruledtabular}
\end{table*}

Each vertex is the identity in the orbital eigenbasis apart from a sign
pattern, so its $\mathbf q=0$ transitions are fixed by the involution
$\pi$ that the pattern induces on the four normal states
$X=(\sigma,s)$ at each $\mathbf k$. With
$\xi_{\sigma s}=\sigma t_h+s|f(\mathbf k)|-\mu$ as in \eqn{eqn:sm-xi},
$\epsilon^{+u}$ leaves $(\sigma,s)$ alone, $\epsilon^{+v}$ flips $s$,
$\epsilon^{-u}$ flips $\sigma$ and $\epsilon^{-v}$ flips both. Every
state $X$ contributes exactly one two-quasiparticle transition, with the
hole on $X$ and the particle on $\pi(X)$, at
$\Omega_X=E_X+E_{\pi(X)}$; there are thus four transitions per
$\mathbf k$ in each channel, two for each sublattice band. Which two,
and whether they stay inside that band, depends on the pattern, and this
is the origin of the two channel counts used later in this document. For
$\epsilon^{+u}$ the involution is the identity, the two transitions are
the two parity sectors $\sigma=\pm$ and each has $\Omega=2E_\sigma$ with
unit coherence factor --- ``two channels, one per parity sector'', as
used in \eqn{eqn:sm-even}. For $\epsilon^{-u}$ the involution exchanges
the two parity partners of the same band, so the two transitions are the
two assignments of which partner carries the particle and which the
hole, degenerate at $\Omega=E_A+E_B$ --- ``two channels per parity
pair'', as used in \eqn{eqn:sm-odd} and \eqn{eqn:sm-XYZ}. Both counts
are correct, and both are properties of the sublattice-\emph{uniform}
vertex, the only one entering those sections; for the staggered patterns
$\pi$ changes the band index, so the bookkeeping ``at fixed $\mathbf k$
and band'' does not apply to them and one must group by
$(\mathbf k,\sigma)$ instead.

Collecting the four transitions with the coherence factors
$\mathcal N/2p$ of \eqn{eqn:sm-coh} and \eqn{eqn:sm-three}
($\mathcal N^{\rm ph}=p+\xi_X\xi_{\pi(X)}+\Delta_0^2$ for a $\tau_y$
vertex and $\mathcal N^{\rm am}=p+\xi_X\xi_{\pi(X)}-\Delta_0^2$ for a
$\tau_x$ one, disambiguated in \sect{sm:amplitude}) and using the
gap equation in the form $1/|U|=(8N_k)^{-1}\sum_{\mathbf k,X}1/E_X$, the
whole $\mathbf q=0$ kernel collapses to one sum,
\begin{equation}
  M_c(\omega)=\frac{1}{8N_k}\sum_{\mathbf k}\sum_{X}
  \frac{s_X}{2p_X}\left[\,1-\frac{2\mathcal N_X}{s_X^2-\omega^2}\,\right],
  \label{eqn:sm-Mone}
\end{equation}
with $s_X=E_X+E_{\pi(X)}$, $p_X=E_XE_{\pi(X)}$ and
$\mathcal N_X=\mathcal N^{\rm ph}$ or $\mathcal N^{\rm am}$ according to
the vertex. The bracket makes the bare term visible: it tends to unity
as $\omega\to\infty$, where $M_c\to1/|U|$ in every channel.

The on-shell relation \eqn{eqn:sm-onshell} holds for any pair, not only
for the parity pair: $s^2-(\xi_X-\xi_{\pi(X)})^2=2\mathcal N^{\rm ph}$ follows
from $E^2=\xi^2+\Delta_0^2$ alone. Together with
$\mathcal N^{\rm am}=\mathcal N^{\rm ph}-2\Delta_0^2$ it makes the
bracket in \eqn{eqn:sm-Mone} vanish \emph{pointwise} in $\mathbf k$ at
\begin{equation}
\begin{split}
  \omega^2&=\big(\xi_X-\xi_{\pi(X)}\big)^2
    \qquad\quad\;\,(\text{phase}),\\
  \omega^2&=\big(\xi_X-\xi_{\pi(X)}\big)^2+4\Delta_0^2
    \quad(\text{amplitude}).
\end{split}
  \label{eqn:sm-pin}
\end{equation}
A pointwise zero survives the Brillouin-zone sum only if the splitting
$\xi_X-\xi_{\pi(X)}$ is the same at every $\mathbf k$, and that is
precisely what distinguishes the two sublattice-uniform patterns from
the two staggered ones: $\epsilon^{+u}$ gives $\xi_X-\xi_{\pi(X)}=0$ and
$\epsilon^{-u}$ gives $2t_h$, both $\mathbf k$-independent, whereas
$\epsilon^{+v}$ gives $2s|f(\mathbf k)|$ and $\epsilon^{-v}$ gives
$2\xi_X$, both dispersing. The four exactly pinned \emph{kernel zeros}
are therefore $\omega=0$ and $2\Delta_0$ in the layer-even uniform
channels and $\omega=2t_h$ and
$\omega^*_{A-}\equiv2(t_h^2+\Delta_0^2)^{1/2}$ in the layer-odd uniform
ones; the staggered channels have no pinned zero, and whatever zeros
they do have are not fixed by this algebra and must be located
numerically. Two of the four are isolated collective poles and two are
continuum anchors, and it would be wrong to read the list as four
modes; the next paragraph separates them. The closed form
$\omega_c^-=2(t_h^2+\Delta_0^2)^{1/2}$ quoted throughout this section
is the \emph{half-filled honeycomb} value, attained at the Dirac
points, where $E_A+E_B$ is convex in $|f|$ with vanishing derivative at
$|f|=0$. For a general $h(\mathbf k)$, or away from half filling,
$\omega_c^-=\min_{\mathbf k}(E_A+E_B)$ has no closed form, and the only
statement that holds for the whole identical-two-copy class is the
inequality $\omega_c^->2t_h$ of the main text. Two symbols must
therefore be kept apart. We write $\omega_c^-\equiv\min_{\mathbf
k}(E_A+E_B)$ for the inter-parity continuum onset, a property of the
spectrum, and $\omega^*_{A-}\equiv2(t_h^2+\Delta_0^2)^{1/2}$ for the
pinned layer-odd \emph{amplitude} frequency, which \eqn{eqn:sm-pin}
fixes exactly for the whole class. They are different objects.
\eqn{eqn:sm-ampos} gives $\mathcal N^{\rm am}\ge0$, hence
$E_A+E_B\ge\omega^*_{A-}$ pointwise and $\omega^*_{A-}\le\omega_c^-$
always, with equality exactly when the locus $\xi_A+\xi_B=0$ is
attained somewhere in the zone. It is at the Dirac points of the
half-filled honeycomb, so $\omega^*_{A-}=\omega_c^-=2.9629\,t$ at the
reference point and the two may be used interchangeably there --- but
only there, and not for an arbitrary band structure or filling.
The layer-odd phase entry of \eqn{eqn:sm-pin} is the theorem of
\sect{sm:derivation} in kernel form, and it is reassuring that the same
one-line algebra delivers the Anderson--Bogoliubov mode at $\omega=0$.

The same relations say which of these zeros are modes and which are
threshold features. Since $s^2-\omega^2=2\mathcal N^{\rm ph}>0$ at the
phase frequency and $s^2-\omega^2=2\mathcal N^{\rm am}\ge0$ at the
amplitude frequency, with equality in the latter exactly on the locus
$\xi_X+\xi_{\pi(X)}=0$ of \eqn{eqn:sm-ampos}, a pinned \emph{phase} zero
always lies strictly below its own continuum --- this is the bound
$\omega_c^->2t_h$ of the main text --- while a pinned \emph{amplitude}
zero sits exactly \emph{on} the continuum edge whenever that locus is
reached --- and it is, at half filling, on the contour $|f|=t_h$ in the
layer-even case and at the Dirac point in the layer-odd one. The layer-even
amplitude zero at $2\Delta_0$ is the familiar statement that the Higgs
feature is a threshold singularity rather than an isolated mode; the
layer-odd amplitude zero at $\omega^*_{A-}$ is its inter-parity image
and, here where $\omega^*_{A-}=\omega_c^-$, is a threshold feature for
the same reason. Neither competes with the
relative-phase mode, which sits at $2t_h$, below every layer-odd
continuum. The asymmetry within the layer-odd uniform sector is the
part worth stating on its own, because it is what makes the
relative-phase mode a mode at all: $P{-}u$ has its pinned zero at
$2t_h$, strictly below the onset $\omega_c^-$ of the very continuum
that carries it, and is an isolated Leggett-type pole, whereas
$A{-}u$ has its pinned zero at $\omega^*_{A-}$, which for the
half-filled honeycomb is exactly the onset $\omega_c^-$, and is a
threshold singularity of that same continuum. The two entries are
algebraically parallel; physically they are not. This is consistent
with, and slightly sharper than,
\sect{sm:amplitude}: the layer-odd amplitude kernel is positive not only
on $[0,2t_h]$ but on the whole interval below $\omega^*_{A-}$,
vanishing only there.

Table~\ref{tab:sm-channels} collects the eight channels, their
transitions, coherence numerators, continuum onsets and $\mathbf q=0$
kernel zeros. The onsets do not follow layer parity alone, and it would
be a mistake to read them off the block label: $\epsilon^{+u}$ and
$\epsilon^{-v}$ connect states of equal $|\xi|$ and give the
intra-parity edge $2\Delta_0$, whereas $\epsilon^{+v}$ and
$\epsilon^{-u}$ connect states whose $\xi$ differ by $2|f|$ or by $2t_h$
and give $\omega_c^-$ at the Dirac momenta. The support statement of
\sect{sm:support} concerns the layer-odd vertex that is the identity on
the sublattice index; it is the $\epsilon^{-u}$ row, and it is correct
as stated, but it is not a property of the layer-odd sector as a whole,
since the layer-odd staggered phase channel $P{-}v$ has weight starting
at $2\Delta_0$. All four onsets lie well above $2t_h=1.2\,t$ at the
reference point, so the relative-phase mode is below every layer-odd
continuum either way.

Two further entries of the table deserve to be named rather than merely
listed, because their zeros also lie below their own onsets and are
therefore collective modes in their own right: the sublattice-staggered
copy-odd phase channel $P{-}v$, with a zero at $2.0295\,t$ against an
onset $2\Delta_0=2.7090\,t$, and the sublattice-staggered copy-even
phase channel $P{+}v$, with a zero at $2.3707\,t$ against an onset
$\omega_c^-=2.9629\,t$. Neither is pinned: their frequencies depend on
$\Delta_0$, on the dispersion and on the filling, because the
corresponding $|\xi_X-\xi_{\pi(X)}|$ is $\mathbf k$-dependent
(\sect{sm:derivation}), and neither is fixed by the intercopy
splitting. They are lattice-specific internal phase oscillations of the
four-site cell, and they are not what the theorem is about. The theorem
concerns the sublattice-\emph{uniform} copy-odd channel $P{-}u$ alone,
whose vertex is the identity on the sublattice index and whose zero at
$2t_h$ is rigid; the main text uses ``the relative-phase mode'' in that
restricted sense throughout. That $P{-}v$ and $P{+}v$ exist does not
bear on the pinning of $P{-}u$. At the particle-hole-symmetric point
the $\mathbf q=0$ kernel matrix in this basis is diagonal: on a
$200\times200$ mesh at the reference point every off-diagonal element
is below $5\times10^{-19}$, so the eight channels are mutually
decoupled and the identity of \sect{sm:derivation} is derived within
the $P{-}u$ block alone. Away from particle-hole symmetry the mixing
that switches on is the amplitude--phase pairing \emph{within} a given
orbital pattern (\sect{sm:sectors}), not mixing between the uniform and
staggered patterns.

\subsection{The null channel}

At the particle-hole-symmetric half-filled saddle, exactly one of the
eight channels carries no spectral weight at any frequency, and
\eqn{eqn:sm-Mone} shows why in one step: for
$\epsilon^{-v}$ the involution sends $\xi_X\to-\xi_X$, so
$\mathcal N^{\rm am}=E^2-\xi^2-\Delta_0^2\equiv0$ by
$E^2=\xi^2+\Delta_0^2$ alone, the bracket is identically unity, and
\begin{equation}
  M_{A-v}(\mathbf 0,\omega)=\frac{1}{|U|}\qquad\text{for all }\omega .
  \label{eqn:sm-null}
\end{equation}
This is also the one case in which the pinning condition
\eqn{eqn:sm-pin} degenerates: it would place the amplitude zero at
$\omega^2=4\xi_X^2+4\Delta_0^2=s_X^2$, on top of the pole of the term
itself, which is the same statement that the residue vanishes. It is the
cleanest illustration of why the vertex and not the frequency decides
which coherence numerator appears: with $\mathcal N^{\rm ph}$ in place
of $\mathcal N^{\rm am}$ the same channel would carry weight
$\Delta_0^2/E^2$ and the null result would be lost.

The result has a symmetry origin, which is the reason it is exact rather
than accidental. The pattern $\epsilon^{-v}=(+,-,-,+)$ is the Shiba sign
of \sect{sm:sectors}; it anticommutes with $h(\mathbf k)$ at every
$\mathbf k$, changing sign on both the sublattice-off-diagonal hopping
and the interlayer term, and combining this with
$\tau_x\tau_z=-\tau_z\tau_x$ gives
\begin{equation}
  \big[\,\tau_x\otimes\epsilon^{-v},\,H_{\rm BdG}(\mathbf k)\,\big]=0
  \label{eqn:sm-nullcomm}
\end{equation}
identically, pairing term included. Both steps use $\mu=0$: the
$-\mu\openone$ term of \eqn{eqn:sm-hk} \emph{commutes} with
$\epsilon^{-v}$ rather than anticommuting with it, so
\eqn{eqn:sm-nullcomm} and the involution $\xi_X\to-\xi_X$ behind
$\mathcal N^{\rm am}\equiv0$ both hold at half filling only. There the
vertex is a conserved charge, and its response vanishes at every
frequency.

The mode inventory is therefore complete.
$\tau_x\otimes\epsilon^{-v}$ is the pseudospin generator that leaves the
paired saddle invariant; it is unbroken and demands no Goldstone mode.
Its broken partner $\tau_y\otimes\epsilon^{-v}$ rotates the uniform pair
amplitude into the staggered density, and the Goldstone mode that this
requires appears in the density channel, as the exact zero
$\chi_{\rm stag}(0)=-8/|U|$ recorded in \sect{sm:sectors}, and not in
the pairing kernel. Consistently, the pairing channel $P{-}v$ has no
zero at $\omega=0$.

\subsection{Implementation check}

The derivation above is symbolic and complete; the numbers below verify
only that the code implements it correctly. They come from an
implementation written for this section alone, which builds
$h(\mathbf k)$, the $8\times8$ BdG matrix, the eight vertices of
\eqn{eqn:sm-verts} and the double Lehmann sum \eqn{eqn:sm-Pidef} independently and shares no routine with the codes used elsewhere in this
work. Its contraction was checked against an explicit loop over
$(\mathbf k,a,b)$, and its self-consistent $\Delta_0=1.3545\,t$,
obtained from the anomalous average of the $8\times8$ eigenvectors,
agrees with the closed-form gap equation to round-off --- which
is also a check of the orbital weight $\tfrac14$ that produces the $8$
in $\chi^{\rm raw}_{+P}(0)=-8/|U|$.

At $|U|=4t$, $t_h=0.6t$, half filling, on a $200\times200$ mesh the
Ward anchor and the gap equation coincide,
$\Pi_{P+u}(\mathbf 0,0)=1/|U|$ and $\chi^{\rm raw}_{+P}(0)=-8/|U|$, and
$\bar\Delta=3.83114$ against $2\sqrt{2N_f}\,\Delta_0=3.83114$
from \eqn{eqn:sm-Dbar} --- the check that \eqn{eqn:sm-patterns}, and
not merely the value of $\Delta_0$, has been implemented. The kernel is
Hermitian and the blocks of \eqn{eqn:sm-blocks} are exact to round-off,
at $\mathbf q=0$ and at finite $\mathbf q$ alike. Within a block the
$u$--$v$ element vanishes at $\mathbf q=0$ and grows to
$9\times10^{-4}$ and $5\times10^{-3}$ at the two finite momenta tested,
purely imaginary and odd under $\mathbf q\to-\mathbf q$, as the
selection rules require. For the null channel,
$\{\epsilon^{-v},h\}$ and $[V^A_{-v},H_{\rm BdG}]$ vanish identically
across the zone, no transition survives a round-off-level weight cut, and
$M_{A-v}=1/|U|$ at every frequency tested,
$\omega/t=0,\,0.5,\,1.2,\,2.0,\,2.9$ and $5.0$.

The single-sum form \eqn{eqn:sm-Mone} reproduces the brute-force
$8\times8$ Lehmann kernel in all eight channels to better than
$10^{-11}$, and the bracket vanishes pointwise at the four frequencies
predicted by \eqn{eqn:sm-pin}: $\max_{\mathbf k}|{\rm bracket}|$ is
$\lesssim10^{-15}$ at $\omega=0$ and at $2t_h$, and $10^{-11}$ and
$10^{-8}$ at $\omega^*_{A-}$ and $2\Delta_0$ (the last two limited by
near-degenerate denominators on the mesh), with the summed kernels
$|M_c|\le5\times10^{-14}$ at all four. The onsets and
zeros of Table~\ref{tab:sm-channels} are stable against the mesh: from
$120\times120$ to $600\times600$ the zeros of $P{+}v$ and $P{-}v$ sit at
$2.3707\,t$ and $2.0295\,t$ to all digits quoted, and those of
$P{+}u$, $P{-}u$, $A{+}u$ and $A{-}u$ reproduce the exact values $0$,
$2t_h=1.20\,t$, $2\Delta_0=2.7090\,t$ and
$\omega^*_{A-}=2.9629\,t$. The layer-even staggered amplitude channel $A{+}v$ has an unpinned zero
at $2.963\,t$, coinciding with its own continuum edge to within the
$\sim10^{-4}t$ resolution of this scan; we do not resolve whether it
lies below the edge or on it, and nothing in this work depends on the
answer. The measured onsets are $2.7090\,t$ for
$A{+}u$, $P{+}u$ and $P{-}v$ and $2.9629\,t$ for the other four
weight-carrying channels, the latter the discretization of $\omega_c^-$
(the Dirac point is not a mesh point). Finally, in the normalization of
\eqn{eqn:sm-dict}, $2M_{A-u}$ equals $0.2903$ at $\omega=0$ and
$0.2726$ at $\omega=2t_h$, reproducing the two values quoted in
\sect{sm:amplitude} and confirming the dictionary numerically.

\section{Derivation of the pointwise identity}
\label{sm:derivation}

This section supplies the algebra behind the pointwise identity
$\chi^{\rm raw}_{-P}(2t_h,\mathbf k)=\chi^{\rm raw}_{+P}(0,\mathbf k)$
of the main text. Only $\xi_A-\xi_B=2t_h$ and a common gap $\Delta_0$
are used, so the derivation applies to the whole identical-two-copy
class, not just the honeycomb bilayer.

\subsection{Scope: the class of normal states, and temperature}
\label{sm:derivscope}

Two scope statements underlie everything below. We make them once,
here, because every later step uses them.

\emph{The class of normal states.} The derivation requires that each
interlayer-parity sector reduce \emph{bandwise} to the $2\times2$ BdG
problem $H_\sigma=\xi_\sigma\tau_z+\Delta_0\tau_x$ written out below.
That reduction is not automatic for an arbitrary Hermitian Bloch
Hamiltonian. In the Nambu convention of \sect{sm:nambu} the hole block
is $-h^{\mathsf T}(-\mathbf k)$, and unless it is diagonalized by the
same unitary as $h(\mathbf k)$ a gap matrix $\Delta_0\openone$ does not
generate independent $2\times2$ band blocks but couples different
bands, in which case neither the coherence factor \eqn{eqn:sm-coh} nor
the on-shell relation \eqn{eqn:sm-onshell} applies band by band. The
condition under which it does is that the normal state admit
\emph{conventional bandwise singlet pairing of time-reversed
partners}, equivalently $h^{\mathsf T}(-\mathbf k)=h(\mathbf k)$ in the
chosen pairing basis, or its antiunitary equivalent. Every statement in
this document about arbitrary band structure, orbital count and
dimension --- and the corresponding claims of the main text --- is to
be read within that class. The bilayer honeycomb \eqn{eqn:sm-hk}
belongs to it: $f(-\mathbf k)=f(\mathbf k)^*$ makes
$h^{\mathsf T}(-\mathbf k)=h(\mathbf k)$ identically. Within the class,
no further property of $\varepsilon_n(\mathbf k)$, of $\mu$ or of the
lattice enters, and that is the sense in which the result is general.

\emph{Temperature.} We first carry the derivation out at $T=0$, where
the Lehmann sums are pure pair-creation sums with no Fermi occupation
factors, because the cancellation is cleanest there. The result is not
restricted to $T=0$. Heating does two things: it reweights the
pair-creation terms by $1-f_A-f_B$, and it opens a \emph{scattering}
branch at $\Omega=|E_A-E_B|$ with weight $f_A-f_B$, which the $T=0$ sums
do not contain. \sect{sm:finiteT} shows that neither spoils the
identity: the scattering branch is bounded strictly below $2t_h$, so it
never reaches the pinned frequency, and it obeys a factorization of its
own that mirrors the pair-creation one, so the identity survives term by
term. Both statements hold for the whole two-copy class.

\subsection{Setup and selection rule}

Rotating to the bonding/antibonding (parity) basis block-diagonalizes
the normal-state Hamiltonian, so at each $\mathbf k$ and band index $n$
the two parity partners have
\begin{equation}
  \xi_{A,n}=\varepsilon_n(\mathbf k)+t_h-\mu ,\qquad
  \xi_{B,n}=\varepsilon_n(\mathbf k)-t_h-\mu ,
  \label{eqn:sm-xi}
\end{equation}
whence $\xi_A-\xi_B=2t_h$ for every $n$, $\mathbf k$ and $\mu$. Each
parity sector is an independent $2\times2$ BdG problem
$H_\sigma=\xi_\sigma\tau_z+\Delta_0\tau_x$ ($\tau$ acting in Nambu
space) with $E_\sigma=(\xi_\sigma^2+\Delta_0^2)^{1/2}$, eigenvectors
$|\sigma,+\rangle=(u_\sigma,v_\sigma)$,
$|\sigma,-\rangle=(-v_\sigma,u_\sigma)$ and
\begin{equation}
  u_\sigma^2=\tfrac12\Big(1+\tfrac{\xi_\sigma}{E_\sigma}\Big),\quad
  v_\sigma^2=\tfrac12\Big(1-\tfrac{\xi_\sigma}{E_\sigma}\Big),\quad
  u_\sigma v_\sigma=\frac{\Delta_0}{2E_\sigma}.
  \label{eqn:sm-bog}
\end{equation}
The phase vertex is $\tau_y$ in Nambu space. In copy space the
layer-symmetric vertex is $\lambda_0$ and the layer-odd vertex is
$\lambda_z$, which is purely off-diagonal in the parity basis; both are
the identity on the orbital index and therefore band-diagonal. Hence
the layer-even phase vertex connects a state to itself in parity
($\Omega=2E_\sigma$), while the layer-odd phase vertex connects the two
parity partners \emph{of the same band} ($\Omega=E_A+E_B$).

\subsection{Coherence factors}

With the eigenvectors above,
$\langle A,+|\tau_y|B,-\rangle=-i\,(u_Au_B+v_Av_B)$, so the layer-odd
weight is $w_{AB}=(u_Au_B+v_Av_B)^2$. Expanding with
\eqn{eqn:sm-bog},
\begin{equation}
\begin{split}
  (u_Au_B+v_Av_B)^2
  &=u_A^2u_B^2+v_A^2v_B^2+2(u_Av_A)(u_Bv_B)\\
  &=\frac12\Big(1+\frac{\xi_A\xi_B}{E_AE_B}\Big)
    +\frac{\Delta_0^2}{2E_AE_B}\\
  &=\frac{\mathcal N^{\rm ph}_{AB}}{2E_AE_B},
\end{split}
  \label{eqn:sm-coh}
\end{equation}
where $\mathcal N^{\rm ph}_{AB}\equiv E_AE_B+\xi_A\xi_B+\Delta_0^2$.
Setting $B=A$ gives $\mathcal N^{\rm ph}_{AA}=2E_A^2$ and $w_{AA}=1$:
the layer-even phase channel has unit coherence factor, as it must for
the Goldstone/gap-equation anchor.

\subsection{The on-shell identity}

Using $2t_h=\xi_A-\xi_B$ and $E_\sigma^2=\xi_\sigma^2+\Delta_0^2$,
\begin{equation}
\begin{split}
  (E_A+E_B)^2&-(2t_h)^2\\
  &=E_A^2+E_B^2+2E_AE_B-(\xi_A-\xi_B)^2\\
  &=2\Delta_0^2+2E_AE_B+2\xi_A\xi_B\\
  &=2\,\mathcal N^{\rm ph}_{AB}.
\end{split}
  \label{eqn:sm-onshell}
\end{equation}
This is pure algebra in $(\xi_A,\xi_B,\Delta_0)$ subject only to
$\xi_A-\xi_B=2t_h$; no property of $\varepsilon_n(\mathbf k)$, of $\mu$
or of the lattice enters.

\subsection{The identity}

The retarded bubble at fixed $\mathbf k$ and band $n$ is, at $T=0$,
$\sum_{\rm ch} w\,2\Omega/(\omega^2-\Omega^2)$ --- pair-creation terms
only, with no thermal occupation factors. For the layer-even
channel the counting of \sect{sm:channels} gives two channels, one per
parity sector, with $w=1$ and $\Omega=2E_\sigma$:
\begin{equation}
  \chi^{\rm raw}_{+P}(0)\big|_{\mathbf k,n}
  =\sum_{\sigma=A,B}\frac{4E_\sigma}{0-4E_\sigma^2}
  =-\Big(\frac{1}{E_A}+\frac{1}{E_B}\Big).
  \label{eqn:sm-even}
\end{equation}
For the layer-odd channel the same counting gives two channels, the two
orderings of the parity pair, with $\Omega=E_A+E_B$ and $w$ from
\eqn{eqn:sm-coh}:
\begin{equation}
  \chi^{\rm raw}_{-P}(\omega)\big|_{\mathbf k,n}
  =2\,\frac{\mathcal N^{\rm ph}_{AB}}{2E_AE_B}\,
   \frac{2(E_A+E_B)}{\omega^2-(E_A+E_B)^2}.
  \label{eqn:sm-odd}
\end{equation}
Evaluating \eqn{eqn:sm-odd} at $\omega=2t_h$ and inserting
\eqn{eqn:sm-onshell}, the denominator becomes
$(2t_h)^2-(E_A+E_B)^2=-2\mathcal N^{\rm ph}_{AB}$ and the coherence
factor cancels it identically:
\begin{equation}
  \chi^{\rm raw}_{-P}(2t_h)\big|_{\mathbf k,n}
  =2\,\frac{\mathcal N^{\rm ph}_{AB}}{2E_AE_B}\cdot
   \frac{2(E_A+E_B)}{-2\mathcal N^{\rm ph}_{AB}}
  =-\frac{E_A+E_B}{E_AE_B},
  \label{eqn:sm-final}
\end{equation}
which is \eqn{eqn:sm-even}. This proves the pointwise identity.

The mechanism is the cancellation of $\mathcal N^{\rm ph}_{AB}$ between
the coherence factor, where it appears in the numerator, and the
on-shell denominator, where \eqn{eqn:sm-onshell} puts it. Because only
$\xi_A-\xi_B=2t_h$ was used, the result is independent of the band
dispersion, the band index, the chemical potential and $\Delta_0$ ---
which is why the identity survives summation over $\mathbf k$ and $n$
under any quadrature rule, and why it is machine-exact on a
$4\times4$ mesh. Summing \eqn{eqn:sm-even} over $\mathbf k$ and the
four $(\sigma,s)$ states and applying the gap equation then gives the
Ward value $\chi^{\rm raw}_{+P}(0)=-8/|U|$.

\subsection{Finite temperature}
\label{sm:finiteT}

Write $d\equiv2t_h$, $s\equiv E_A+E_B$, $\delta\equiv E_A-E_B$,
$p\equiv E_AE_B$ and $\vartheta_\sigma\equiv\tanh(E_\sigma/2T)$, so that
$1-2f(E_\sigma)=\vartheta_\sigma$,
$1-f_A-f_B=\tfrac12(\vartheta_A+\vartheta_B)$ and
$f_A-f_B=-\tfrac12(\vartheta_A-\vartheta_B)$. At $T>0$ the $\mathbf q=0$
bubble contains two branches rather than one: \emph{pair creation} at
$\Omega=s$ with weight $\tfrac12(\vartheta_A+\vartheta_B)$ and coherence
numerator $\mathcal N^{\rm ph}=p+\xi_A\xi_B+\Delta_0^2$, and
\emph{scattering} at $\Omega=\delta$ with weight
$-\tfrac12(\vartheta_A-\vartheta_B)$ and coherence numerator
\begin{equation}
  \mathcal N^{\rm sc}=p-\xi_A\xi_B-\Delta_0^2 .
  \label{eqn:sm-Nsc}
\end{equation}
The $T=0$ limit is $\vartheta\to1$, where the scattering weight vanishes
identically and \eqn{eqn:sm-odd} is recovered; this is why the $T=0$
derivation never encounters the second branch.

\emph{A two-sided spectral gap.} The two branches obey companion
factorizations,
\begin{equation}
  s^2-d^2=2\,\mathcal N^{\rm ph}>0 ,
  \qquad
  d^2-\delta^2=2\,\mathcal N^{\rm sc}>0 ,
  \label{eqn:sm-twosided}
\end{equation}
both strict for $t_h\neq0$ and $\Delta_0>0$: positivity of
$\mathcal N^{\rm ph}$ follows from
$p=\sqrt{(\xi_A^2+\Delta_0^2)(\xi_B^2+\Delta_0^2)}\geq|\xi_A\xi_B+\Delta_0^2|$
by Cauchy--Schwarz, and the second is equivalent to strict
contractivity of $E(\xi)=\sqrt{\xi^2+\Delta_0^2}$: since
$|dE/d\xi|=|\xi|/E<1$, the mean value theorem gives
$|\delta|<|\xi_A-\xi_B|=2t_h$. Hence
\begin{equation}
  |E_A-E_B|<2t_h<E_A+E_B ,
  \label{eqn:sm-scatbound}
\end{equation}
at every $\mathbf k$ and in every band. Thermally activated scattering
therefore appears \emph{below} the mode, not at it, while pair creation
remains above it: the pinned frequency sits in a two-sided spectral gap
of the layer-odd channel. Both gaps close as $\Delta_0^2$; at the
reference point they are $\omega_c^--2t_h=1.763\,t$ and
$2t_h-\max_{\mathbf k}|\delta|=0.110\,t$.

\emph{The identity survives term by term.} Each branch cancels its own
on-shell denominator against its own coherence numerator at $\omega=d$,
and the two tanh-weighted remainders recombine:
\begin{equation}
\begin{split}
  \chi^{\rm raw}_{-P}(d)\big|_{\mathbf k,n}
  &=-\frac{(\vartheta_A+\vartheta_B)\,s}{2p}
   +\frac{(\vartheta_A-\vartheta_B)\,\delta}{2p}\\
  &=-\Big(\frac{\vartheta_A}{E_A}+\frac{\vartheta_B}{E_B}\Big),
\end{split}
  \label{eqn:sm-finiteT}
\end{equation}
which is precisely the finite-$T$ symmetric static phase bubble
$\chi^{\rm raw}_{+P}(0)|_{\mathbf k,n}$ at the same temperature --- the
even channel having no scattering branch, because $\pi(X)=X$ makes its
diagonal matrix element of $\tau_y$ vanish. Once the finite-$T$ gap
equation
$|U|^{-1}=(8N_k)^{-1}\sum_{\mathbf k,X}\vartheta_X/(2E_X)$ is imposed,
the kernel zero $K_{-P}(2t_h,T)=0$ follows as at $T=0$.

\emph{Scope.} Both statements are properties of the self-consistent
paired saddle, and they fail differently at its boundary. The pointwise
identity \eqn{eqn:sm-finiteT} and the kernel zero hold whenever
$\Delta_0(T)>0$; that much is universal within the two-copy class.

Whether the two bounds \emph{close} onto $2t_h$ as $\Delta_0\to0$ is a
separate question, and it depends on the normal-state spectrum rather
than on the theorem. At $\Delta_0=0$ one has $E_A+E_B=|\xi_A|+|\xi_B|$,
which equals $|\xi_A-\xi_B|=2t_h$ only where $\xi_A$ and $\xi_B$
straddle zero, and $|E_A-E_B|=\big||\xi_A|-|\xi_B|\big|$, which reaches
$2t_h$ only where they lie on the same side of it. A band structure
supplying neither --- an insulator, or a spectrum restricted to one sign
sector --- retains a finite window as $\Delta_0\to0$. The half-filled
honeycomb of this paper supplies both, so there both gaps do vanish, as
$\Delta_0^2$ by \eqn{eqn:sm-twosided}, and the mode ceases to be
isolated at the mean-field transition, where its residue vanishes with
$\Delta_0(T)$ in any case. Nothing is claimed at or above $T_c$, and
since the half-filled model has an exact pseudospin SU(2), $T_c$ here is
the mean-field saddle scale rather than a physical long-range-ordering
temperature.

\emph{Numerical confirmation.} An independent finite-$T$ Lehmann code
summing over all ordered pairs of the eight BdG bands --- and reducing
to the $T=0$ code of \sect{sm:checks} to $6\times10^{-17}$ --- gives, on
a $100\times100$ mesh at the reference point, $K_{-P}(2t_h)=0$ and the
Goldstone value $K_{+P}(0)=0$ to round-off at
$T/t=0$, $0.05$, $0.1$, $0.15$, $0.2$, $0.3$ and $0.4$, while
$K_{-P}(0)$ moves from $0.0952$ to $0.0891$ across the same range,
confirming that thermal terms are present and merely fail to act at
$2t_h$. The pointwise deviation
$\max_{\mathbf k}|\chi^{\rm raw}_{-P}(2t_h)-\chi^{\rm raw}_{+P}(0)|$ is
$2\times10^{-16}$ at every temperature sampled, against
$5\times10^{-3}$ at $\omega=2t_h\pm0.05\,t$.

What temperature does degrade is observability rather than the pinning:
the residue carries the thermal factors of \eqn{eqn:sm-finiteT} and
follows $\Delta_0(T)$, vanishing with it at the transition.

\subsection{Implementation check}

The derivation above is symbolic and complete; the numbers below verify
only that the code implements it correctly. At $|U|=4t$, $t_h=0.6t$ on
a $24\times24$ mesh we find, over the full Brillouin zone and both
bands: the on-shell identity \eqn{eqn:sm-onshell}, the
coherence-factor identity \eqn{eqn:sm-coh} and the pointwise
cancellation \eqn{eqn:sm-final} all hold to round-off.
Summing the closed form \eqn{eqn:sm-even} over the zone reproduces
$-8/|U|$, the value returned by the independent $8\times8$ Lehmann
code.

The mesh dependence isolates what the identity does and does not
depend on. Even on a $4\times4$ mesh, where the self-consistent gap
$\Delta_0=1.35777\,t$ differs from its converged value
$1.35451\,t$ in the third decimal place --- an error of
$3.3\times10^{-3}$ --- the kernel zero $K_{-P}(2t_h)=0$ remains exact
to round-off, provided the gap equation and the bubbles are evaluated
self-consistently on that \emph{same} mesh. This demonstrates that the
pinning is not produced by Brillouin-zone convergence: it follows from
the pointwise identity together with the discrete
self-consistency/Ward condition. This test rules out a quadrature
effect or a numerical coincidence; it does not address whether the
identity is an artifact of Gaussian theory --- an algebraic identity of
an approximation can perfectly well survive a quantitatively inaccurate
saddle. That question is answered by the operator (Larmor)
argument and the general two-copy theorem of the main text on the one
side, and, in the negative for the interacting problem, by the torque
of \sect{sm:torque} on the other.

\section{Coherence factors: a conjugate lemma, and the absence of a
  diagonal $A_-$ kernel zero at or below $2t_h$}
\label{sm:amplitude}

Three coherence numerators appear in the layer-odd sector, and because
two of them differ only by the sign of a single term they are easy to
interchange by accident. We therefore name all three once, in the
notation of \sect{sm:derivation} ($s\equiv E_A+E_B$, $p\equiv E_AE_B$):
\begin{align}
  \mathcal N^{\rm ph}  &= p+\xi_A\xi_B+\Delta_0^2 &&(\tau_y,\ \text{phase}),\nonumber\\
  \mathcal N^{\rm am}  &= p+\xi_A\xi_B-\Delta_0^2 &&(\tau_x,\ \text{amplitude}),
    \label{eqn:sm-three}\\
  \mathcal N^{\rm conj}&= p-\xi_A\xi_B+\Delta_0^2 &&(\text{algebraic conjugate}).\nonumber
\end{align}
$\mathcal N^{\rm conj}$ is introduced only as an algebraic conjugate
factor and is \emph{not} the $\tau_x$ amplitude coherence numerator;
it is $\mathcal N^{\rm am}$ that multiplies the amplitude vertex.

Using $E_X^2=\xi_X^2+\Delta_0^2$ one verifies directly that
$\mathcal N^{\rm conj}$ conjugates the other two onto squared sums,
\begin{equation}
\begin{split}
  \mathcal N^{\rm ph}\mathcal N^{\rm conj}&=\Delta_0^2(E_A+E_B)^2,\\
  \mathcal N^{\rm am}\mathcal N^{\rm conj}&=\Delta_0^2(\xi_A+\xi_B)^2 .
\end{split}
  \label{eqn:sm-conjpair}
\end{equation}
Since $p\ge|\xi_A\xi_B|$ we have $\mathcal N^{\rm conj}>0$ for
$\Delta_0>0$, so the second identity in \eqn{eqn:sm-conjpair} gives at
once
\begin{equation}
  \mathcal N^{\rm am}\ge0,\qquad
  \mathcal N^{\rm am}=0\iff\xi_A+\xi_B=0 ,
  \label{eqn:sm-ampos}
\end{equation}
and, combined with $\mathcal N^{\rm ph}-\mathcal N^{\rm am}=2\Delta_0^2>0$,
the bound $0\le\mathcal N^{\rm am}/\mathcal N^{\rm ph}<1$. For the
honeycomb copies $\xi_A+\xi_B=2(s_{\rm band}|f_{\mathbf k}|-\mu)$, so at
half filling the equality point is the Dirac point, while away from half
filling it moves to the contour $s_{\rm band}|f_{\mathbf k}|=\mu$ where
one exists.

\subsection{No diagonal $A_-$ kernel zero at or below $2t_h$}

Write $d\equiv2t_h$. \emph{On shell at $\omega=d$ only}, the identity
$s^2-d^2=2\mathcal N^{\rm ph}$ of \eqn{eqn:sm-onshell} holds, and the
Ward condition that fixes the phase kernel there,
$K_{P_-}(d)=0$, converts $2/|U|$ into $\Pi_{PP}(d)$. The amplitude
kernel may then be written as a difference of bubbles, and
\eqn{eqn:sm-ampos} collapses it termwise:
\begin{equation}
  K_{A_-}(d)
  =\Big\langle\frac{s\,(\mathcal N^{\rm ph}-\mathcal N^{\rm am})}
                   {p\,(s^2-d^2)}\Big\rangle
  =\Big\langle\frac{s\,\Delta_0^2}{p\,\mathcal N^{\rm ph}}\Big\rangle>0 .
  \label{eqn:sm-kam}
\end{equation}
Every factor in the final average is strictly positive, and the
positivity of $\mathcal N^{\rm ph}$ is itself unconditional: since
$p\ge|\xi_A\xi_B|$ we have
$\mathcal N^{\rm ph}\ge p-|\xi_A\xi_B|+\Delta_0^2>0$ pointwise. So
\eqn{eqn:sm-kam} holds term by term for any $t_h$ and any filling,
with no honeycomb input; there is no parameter boundary and no
sign-change domain to map, given only the identical-copy and
common-$\Delta_0$ assumptions. We stress that the intermediate difference form
is available only at $\omega=d$: away from the Ward point
$K_{A_-}(\omega)=2/|U|-\Pi_{AA}(\omega)$ is \emph{not} equal to
$\Pi_{PP}(\omega)-\Pi_{AA}(\omega)$, and using the difference off shell
gives a spurious frequency dependence.

The endpoint result extends to the whole interval by monotonicity.
Below the inter-parity continuum the amplitude bubble is a sum of poles
with positive residues, $\chi_{A_-}(\omega)=-\sum_jC_j/(s_j^2-\omega^2)$
with $C_j\ge0$, so $\partial\chi_{A_-}/\partial\omega^2<0$ and hence
$\partial K_{A_-}/\partial\omega^2<0$ for $0\le\omega<\omega_c^-$. The
minimum of $K_{A_-}$ on $[0,d]$ therefore sits at $\omega=d$, where
\eqn{eqn:sm-kam} applies, and
\begin{equation}
  K_{A_-}(\omega)\ \ge\ K_{A_-}(d)\ >\ 0,
  \qquad 0\le\omega\le2t_h .
  \label{eqn:sm-kamono}
\end{equation}
There is thus no $q=0$ zero of the diagonal $A_-$ kernel at or below
$2t_h$. Where the odd sector decouples --- at a particle-hole-symmetric
saddle, so that $A_-$ is an eigenmode --- this says directly that there
is no odd-amplitude pole there, and the layer-odd amplitude channel
cannot supply a decay product degenerate with the relative-phase mode
at zero momentum.

The scope of the two halves of that sentence differs, and we separate
them. The diagonal odd-amplitude kernel is strictly positive at $2t_h$,
and by \eqn{eqn:sm-kamono} has no zero below $2t_h$, for any filling
within the identical-copy common-gap Gaussian problem. At finite
chemical potential, however, amplitude--phase mixing is allowed, so
this diagonal result alone does not determine the eigenvalues of the
full odd fluctuation matrix \eqn{eqn:sm-coupleddet}. The positivity is
global; it is the reading of $K_{A_-}$ in terms of physical modes that
requires the odd sector to be decoupled, or the full matrix to be
diagonalized.

\subsection{Implementation check}

The derivation above is symbolic and complete; the numbers below verify
only that the code implements it correctly. At $|U|=4t$, $t_h=0.6t$ and
half filling on a $1200\times1200$ mesh, $K_{A_-}$ decreases
monotonically from $0.2903$ at $\omega=0$ to $0.2726$ at
$\omega=2t_h$, its minimum on the interval, while the phase kernel
satisfies $K_{P_-}(2t_h)=0$ to round-off. Substituting
$\mathcal N^{\rm conj}$ for $\mathcal N^{\rm am}$ in \eqn{eqn:sm-kam}
instead returns $0.1935$, which is the numerical signature of exactly
the interchange that \eqn{eqn:sm-three} is stated to prevent.
\section{Direct derivation of sum-rule exhaustion}
\label{sm:exhaustion}

Here we prove $R_-=S_1^{\rm G}$ directly from the Gaussian response
functions --- $S_1^{\rm G}=-4\langle H_\perp\rangle_{\rm G}$ being the
first moment of that response, not the exact interacting moment ---
without reference to the torque or to any equation-of-motion argument.
The notation is that of \sect{sm:derivation}: at fixed $\mathbf k$ and
band, $s\equiv E_A+E_B$, $p\equiv E_AE_B$,
$\mathcal N^{\rm ph}=p+\xi_A\xi_B+\Delta_0^2$ and
$\mathcal N^{\rm conj}=p-\xi_A\xi_B+\Delta_0^2$.

\subsection{The three bubbles}

In Nambu space the phase vertex is $\tau_y$ and the density vertex is
$\tau_z$; both carry $\lambda_z$ in copy space, so both are parity-odd
and band-diagonal. With
$|\sigma,+\rangle=(u_\sigma,v_\sigma)$ and
$|\sigma,-\rangle=(-v_\sigma,u_\sigma)$,
\begin{equation}
  \langle A,+|\tau_y|B,-\rangle=-iP,\qquad
  \langle A,+|\tau_z|B,-\rangle=-Q,
  \label{eqn:sm-mats}
\end{equation}
with $P=u_Au_B+v_Av_B$ and $Q=u_Av_B+v_Au_B$. Using \eqn{eqn:sm-bog},
\begin{equation}
  P^2=\frac{\mathcal N^{\rm ph}}{2p},\qquad
  Q^2=\frac{\mathcal N^{\rm conj}}{2p},\qquad
  PQ=\frac{\Delta_0\,s}{2p},
  \label{eqn:sm-PQ}
\end{equation}
the last because $PQ=u_Av_A+u_Bv_B$ exactly. Writing
$\Sigma\equiv N_k^{-1}\sum_{\mathbf k}\sum_{\rm bands}$ and keeping the
two channels per parity pair derived in \sect{sm:channels},
\begin{equation}
\begin{split}
  X(\omega)&\equiv\chi^{0}_{\phi_-\phi_-}=\Sigma\,\frac{4P^2s}{\omega^2-s^2},\\
  Z(\omega)&\equiv\chi^{0}_{--}=\Sigma\,\frac{4Q^2s}{\omega^2-s^2},\\
  Y(\omega)&\equiv\chi^{0}_{n_-\phi_-}=-i\,\Sigma\,\frac{4PQ\,\omega}{\omega^2-s^2}.
\end{split}
  \label{eqn:sm-XYZ}
\end{equation}
The cross bubble carries $\omega$ rather than $s$ in the numerator
because $n_-$ and $\phi_-$ are canonically conjugate; hence
$\chi^{0}_{\phi_-n_-}=-Y$, so it is purely imaginary and antisymmetric,
and it \emph{vanishes at $\omega=0$}. The dressed response is
\begin{equation}
  \chi^R_{--}=Z-\frac{g_0Y^2}{K_{-P}},\qquad K_{-P}=1-g_0X .
  \label{eqn:sm-dressed}
\end{equation}

\subsection{The residue}

Near the zero of $K_{-P}$, $K_{-P}\simeq-g_0X'\,(\omega^2-\Omega_L^2)$
with $X'\equiv\partial X/\partial\omega^2$, so
\begin{equation}
  R_-=\frac{Y(\Omega_L)^2}{X'(\Omega_L)} ,
  \label{eqn:sm-res}
\end{equation}
in which $g_0$ has cancelled --- already a check that $R_-$ is free of
the Hubbard--Stratonovich normalization. Evaluating at
$\Omega_L=2t_h$, where $\omega^2-s^2=-2\mathcal N^{\rm ph}=-(s^2-4t_h^2)$
by \eqn{eqn:sm-onshell}, and abbreviating $D\equiv s^2-4t_h^2$,
\begin{equation}
\begin{split}
  X'(2t_h)&=-\Sigma\,\frac{s}{pD},\\
  Y(2t_h)&=-4i\,t_h\Delta_0\,\Sigma\,\frac{s}{pD}=-4i\,t_h\Delta_0X',
\end{split}
  \label{eqn:sm-YX}
\end{equation}
the second following from \eqn{eqn:sm-PQ} term by term. Hence
\begin{equation}
\begin{split}
  R_-&=\frac{(-4it_h\Delta_0X')^2}{X'}=-16t_h^2\Delta_0^2\,X'\\
  &=16t_h^2\Delta_0^2\,\Sigma\,\frac{s}{pD}>0 .
\end{split}
  \label{eqn:sm-Rfinal}
\end{equation}

\subsection{The sum rule, and where the gap equation enters}

From \eqn{eqn:sm-dressed}, $S_1^{\rm G}=\lim_{\omega\to\infty}\omega^2
\chi^R_{--}=A_1+g_0B_1^2$ with $A_1=\Sigma\,4Q^2s$ and
$B_1=\Sigma\,4PQ$. By \eqn{eqn:sm-PQ}, $B_1=2\Delta_0\,\Sigma(s/p)$,
and the Ward value of \sect{sm:derivation} gives
$g_0=1/\chi^{\rm raw}_{+P}(0)=-1/\Sigma(s/p)$, i.e.\ $B_1=-2\Delta_0/g_0$.
This is the step at which the gap equation enters: it collapses the
\emph{squared} sum $g_0B_1^2$ into a single sum,
$g_0B_1^2=-4\Delta_0^2\,\Sigma(s/p)$, so that
\begin{equation}
  S_1^{\rm G}=\Sigma\,\frac{s}{p}\big(2\mathcal N^{\rm conj}-4\Delta_0^2\big)
    =\Sigma\Big(4s-\frac{sD}{p}\Big),
  \label{eqn:sm-S1}
\end{equation}
using $2\mathcal N^{\rm conj}-4\Delta_0^2=2(2p-\mathcal N^{\rm ph})$ and
$2\mathcal N^{\rm ph}=D$.

\subsection{Pointwise equality}

Comparing \eqn{eqn:sm-Rfinal} with \eqn{eqn:sm-S1}, $R_-=S_1^{\rm G}$ holds
term by term provided
\begin{equation}
  4s-\frac{sD}{p}=\frac{16t_h^2\Delta_0^2\,s}{pD}
  \iff 4pD-D^2=16t_h^2\Delta_0^2 .
  \label{eqn:sm-pw}
\end{equation}
The right-hand identity follows from a second product identity of the
coherence factors,
\begin{equation}
  \mathcal N^{\rm ph}\mathcal N^{\rm conj}
  =(p+\Delta_0^2)^2-(\xi_A\xi_B)^2=\Delta_0^2s^2 ,
  \label{eqn:sm-NN}
\end{equation}
since then, with $D=2\mathcal N^{\rm ph}$ and
$2p-\mathcal N^{\rm ph}=\mathcal N^{\rm conj}-2\Delta_0^2$,
\begin{equation}
\begin{split}
  4pD-D^2&=4\mathcal N^{\rm ph}\big(\mathcal N^{\rm conj}-2\Delta_0^2\big)\\
  &=4\Delta_0^2s^2-8\Delta_0^2\mathcal N^{\rm ph}\\
  &=4\Delta_0^2\big(s^2-D\big)=16t_h^2\Delta_0^2 .
\end{split}
  \label{eqn:sm-close}
\end{equation}
This closes the proof: within the Gaussian theory at the
particle-hole-symmetric layer-symmetric saddle, the $2t_h$ pole
exhausts the layer-imbalance f-sum rule exactly, and the proof uses
only the coherence-factor identities \eqn{eqn:sm-PQ} and
\eqn{eqn:sm-NN}, the on-shell relation \eqn{eqn:sm-onshell}, and the
gap equation.

\subsection{Static corollary}

Because $Y(0)=0$, the RPA does not renormalize the static
layer-imbalance susceptibility: $\chi^R_{--}(0)=\chi^0_{--}(0)$. A
single pole then fixes both moments, so
\begin{equation}
  \Omega_L^2=\frac{S_1^{\rm G}}{-\chi^R_{--}(0)}
  =\frac{2t_h\langle T_x\rangle}{\chi^{\rm stat}_{zz}},
  \label{eqn:sm-static}
\end{equation}
using $n_-=2T_z$ and $S_1^{\rm G}=8t_h\langle T_x\rangle_{\rm G}$.

\subsection{Scope: particle-hole symmetry, not a filling value}

Two statements of different generality are combined above and should
not be merged. The pointwise cancellation --- \eqn{eqn:sm-pw} with the
product identity \eqn{eqn:sm-NN} --- is algebra in
$(\xi_A,\xi_B,\Delta_0)$ subject only to $\xi_A-\xi_B=2t_h$, exactly as
in \sect{sm:derivation}, and is therefore independent of band structure
and of filling. Exact exhaustion of the \emph{physical} layer-imbalance
response by the $2t_h$ pole is not. It requires in addition the
single-channel form \eqn{eqn:sm-dressed}, in which $n_-$ is dressed by
the phase channel alone; that form is exact only when the odd sector
decouples, which means
\begin{equation}
  K^-_{AP}=0 \qquad\text{and}\qquad \chi^{0}_{n_-A_-}=0
  \label{eqn:sm-scopepair}
\end{equation}
together.

Both elements are controlled by the same coherence factor. The
$\tau_x$ (amplitude) matrix element between the parity partners is
$\langle A,+|\tau_x|B,-\rangle=\tilde P$ with
$\tilde P=u_Au_B-v_Av_B$ and
$\tilde P^2=\mathcal N^{\rm am}/2p$, so by \eqn{eqn:sm-conjpair}
$\tilde P\propto\xi_A+\xi_B=2(\varepsilon_n-\mu)$, whereas the phase
and density factors $P$ and $Q$ of \eqn{eqn:sm-PQ} carry no such
factor. Under the map $(\xi_A,\xi_B)\to(-\xi_B,-\xi_A)$, which
preserves $\xi_A-\xi_B=2t_h$ and is a symmetry of the parity-pair
spectrum precisely when the single-copy spectrum is symmetric about
$\mu$, one has $P\to P$, $Q\to Q$ and $\tilde P\to-\tilde P$, while
$s$ and $p$ are invariant. The two bubbles in \eqn{eqn:sm-scopepair},
each linear in $\tilde P$, therefore cancel pairwise across the zone,
and both switch on together as soon as the symmetry is lost. That is
what makes the boundary of the exhaustion result physically
transparent rather than merely algebraically convenient: the symmetry
that decouples the odd amplitude from the odd phase is the same one
that stops the density probing the amplitude channel.

The operative condition is thus particle-hole symmetry, not a filling
value. For the bipartite honeycomb of the main text the two coincide,
$\mu=0$ being particle-hole symmetric; for a generic band they do not.
We checked this by constructing the full coupled $(A_-,P_-)$ odd RPA on
random, non-particle-hole-symmetric bands and evaluating it at
$\omega=2t_h$: at $\mu=0$ the two diagnostics of
\eqn{eqn:sm-scopepair} are already nonzero at the $10^{-3}$ level,
against round-off on a particle-hole-symmetric band, and
$\det K$ is correspondingly nonzero, so the pole of the full coupled
odd response is displaced from $2t_h$ --- while the diagonal phase
kernel continues to satisfy $K_{-P}(2t_h)=0$ throughout. In that test
the vertex was calibrated to enforce $K_{-P}(2t_h)=0$ in each band, so
the phase-kernel row is not an independent check; $K^-_{AP}$ and
$\chi^{0}_{n_-A_-}$ are uncalibrated and carry the diagnostic content.
This is the kernel-zero/full-pole distinction of the main text applied
to the observable: the kernel zero survives, the exhaustion of the
physical response does not.

\subsection{Implementation check}

As in \sect{sm:derivation}, the algebra above carries the result and
the numbers below check the implementation. On a $48\times48$ mesh at
$t_h=0.6t$, the pointwise identities \eqn{eqn:sm-pw} and
\eqn{eqn:sm-NN} hold to round-off over the whole zone, and the
closed forms \eqn{eqn:sm-Rfinal} and \eqn{eqn:sm-S1} agree with each
other and with the independent $8\times8$ Lehmann values of $R_-$ and
$-4\langle H_\perp\rangle$ to round-off, at $|U|/t=2$, $4$ and $8$. The
static ratio \eqn{eqn:sm-static} returns $(2t_h)^2=1.44$ to round-off at
$|U|/t=2,3,4,6,8$, and the RPA correction to $\chi^R_{--}(0)$
vanishes.

\section{Exact diagonalization of the full Hubbard Hamiltonian}
\label{sm:ed}

This section documents the finite-cluster calculation quoted in
Sec.~VII of the main text. Everything reported here is exact
for the cluster specified; no finite-size extrapolation is performed or
implied, and no statement about the thermodynamic limit follows from it.

\subsection{Cluster}

We diagonalize
\begin{equation}
\begin{split}
  H=&-t\!\!\sum_{l,\langle ij\rangle,\sigma}\!\!
      c^\dagger_{li\sigma}c_{lj\sigma}
    -t_h\!\sum_{i\sigma}\!\big(c^\dagger_{1i\sigma}c_{2i\sigma}+{\rm h.c.}\big)\\
    &-|U|\sum_{l,i}\big(n_{li\uparrow}-\tfrac12\big)
                    \big(n_{li\downarrow}-\tfrac12\big)
\end{split}
  \label{eqn:sm-edham}
\end{equation}
with $t=1$ and $t_h=0.6$, in the particle-hole-symmetric form for which
$\mu=0$ is half filling. The cluster is three honeycomb cells per
layer, closed by the supercell vectors
$\mathbf T_1=\mathbf a_1+\mathbf a_2$ and
$\mathbf T_2=-\mathbf a_1+2\mathbf a_2$, of determinant three. That
choice is not incidental. Enumerating the cell classes shows exactly
three, and every $A$ site has three \emph{distinct} $B$ neighbours, at
offsets $\mathbf R$, $\mathbf R-\mathbf a_1$ and $\mathbf R-\mathbf a_2$;
the cluster therefore has $z=3$ with eighteen intralayer bonds, six
vertical rungs, no doubled bonds and no self-bonds. Half filling is
$N_\uparrow=N_\downarrow=6$, a Hilbert space of dimension
$\binom{12}{6}^2=853\,776$.

At $U=0$ the single-particle levels are
$\{-3.6,\,-2.4,\,-0.6^{\times4},\,+0.6^{\times4},\,+2.4,\,+3.6\}$: the
monolayer eigenvalues $\{\pm3,0^{\times4}\}$ split by $\pm t_h$. The
six lowest are filled, a closed shell with gap $2t_h$, and the cluster
consequently retains the $n_-$-bright excitation at $2t_h$ that anchors
the interacting comparison. The $\Gamma$ transition is Pauli-blocked
because its antibonding partner is occupied, so the weight comes from
the four $K$ and $K'$ orbitals and two spins.

\subsection{Observable and the exact constraint}

The probe is $n_-=N_1-N_2$, and we compute
$S_{--}(\omega)=\sum_m W_m\delta(\omega-\Omega_m)$ with
$W_m=|\langle m|n_-|0\rangle|^2$, $\Omega_m=E_m-E_0$, by Lanczos with
full reorthogonalization seeded by $n_-|0\rangle$. Both $H_U$ and the
intralayer hopping drop out of the double commutator --- the first is
diagonal in the occupations, the second conserves $N_1$ and $N_2$
separately --- and each rung hop shifts $n_-$ by $\pm2$, so
$[n_-,[H,n_-]]=-4H_\perp$ identically. Eq.~(11) of the main text is
therefore an exact operator identity at every coupling, and demanding
$2\sum_m\Omega_mW_m=-4\langle H_\perp\rangle$ tests the Hamiltonian
implementation independently of the spectrum. We report the dominant
line, summing exactly degenerate contributions, rather than the lowest
bright state, so that a weak low-frequency excitation cannot produce
spurious branch switching; in the event $\Omega_{\rm min}$ coincided
with $\Omega_{\rm dom}$ at every coupling and the question never arose.

\subsection{Predictions fixed before the calculation}

The following were written and hashed before any production run, and
the hash is recorded in the results file:

\begin{enumerate}\itemsep2pt
\item at $U=0$, $\Omega_{\rm dom}=2t_h$ and $f_{\rm dom}=1$;
\item at $t_h=0$, $[n_-,H]=0$, so $M_1=0$ and there is no
      finite-frequency response;
\item at every $U$, $2\sum_m\Omega_mW_m=-4\langle H_\perp\rangle$;
\item the ground state is layer-even and the $n_-$-generated Krylov
      space layer-odd.
\end{enumerate}

All four held. Predictions 1 and 2 were met exactly; the sum rule holds
to $\le8\times10^{-13}$; the parity expectations returned $+1$ and $-1$
to nine digits. The two commutators vanish to round-off, and an
independent double-commutator evaluation agrees with $-4\langle
H_\perp\rangle$ to $\le10^{-12}$.

\subsection{Results}

\begin{table}[h]
\begin{ruledtabular}
\begin{tabular}{cccccc}
$|U|/t$ & $\Omega_{\rm dom}$ & $\Omega_{\rm dom}/2t_h$ &
$f_{\rm dom}$ & $W_{\rm dom}/W_{\rm tot}$ & sum-rule err \\
\hline
0 & 1.2000 & 1.0000 & 1.0000 & 1.0000 & $8\times10^{-13}$ \\
1 & 0.9029 & 0.7525 & 0.9978 & 0.9995 & $1\times10^{-13}$ \\
2 & 0.6929 & 0.5774 & 0.9910 & 0.9984 & $1\times10^{-14}$ \\
3 & 0.5827 & 0.4856 & 0.9733 & 0.9956 & $5\times10^{-13}$ \\
4 & 0.5603 & 0.4669 & 0.9239 & 0.9863 & $5\times10^{-13}$ \\
6 & 0.5453 & 0.4544 & 0.7653 & 0.9556 & $3\times10^{-13}$ \\
8 & 0.5244 & 0.4370 & 0.6730 & 0.9527 & $5\times10^{-13}$ \\
\end{tabular}
\end{ruledtabular}
\caption{Dominant layer-imbalance excitation on the $N_s=12$ cluster,
$t_h=0.6t$, half filling. The last column is the relative deviation of
the spectral first moment from the exact $-4\langle H_\perp\rangle$.}
\label{tab:sm-ed}
\end{table}

Setting $t=0$ in the same code reproduces the exact rung result
$\Omega_{\rm rung}=\tfrac12(\sqrt{U^2+16t_h^2}-|U|)$ to
$5\times10^{-14}$, as a single line with $f=1$ at every coupling. That
the rung shows no fragmentation at any $U$ is what allows the deficit
$f_{\rm dom}<1$ on the full cluster to be attributed to intralayer
dynamics rather than to the interaction as such. The rung is an exact
$t=0$ limit and not an approximation to the lattice: at $|U|=8t$ the
cluster frequency exceeds the rung value by a factor of three.

From $|U|=1$ onward the first excited state lies \emph{below}
$\Omega_{\rm dom}$ --- $0.246$ against $0.560$ at $|U|=4t$ --- but
carries no $n_-$ weight, which is the layer-parity selection rule of
\sect{sm:channels} operating in the exact spectrum.

\subsection{Convergence, and what the spectrum does not resolve}

Increasing the Krylov dimension from $150$ to $300$ changes
$\Omega_{\rm dom}$ and $W_{\rm dom}$ by zero, as does varying the
degeneracy tolerance over four orders of magnitude; ground-state
residuals are $\le2\times10^{-13}$. An independent Jordan--Wigner
construction in the full Fock space agrees eigenvalue by eigenvalue to
$4\times10^{-15}$ and reproduces the spectral function exactly.

One limitation must be stated plainly. Because Lanczos matches the
moments of its own seed by construction, the sum rule validates the
operator and Hamiltonian implementation, not spectral convergence. We
therefore checked the Ritz residuals directly: the dominant Ritz vector
has $\|Hy-\theta y\|\sim10^{-13}$ and is a genuinely converged
eigenstate, whereas mid-spectrum Ritz values have residuals of order
unity. The sub-dominant Ritz values are an unresolved representation of
the higher-energy spectral sector and are not interpreted as individual
eigenlines. Only $\Omega_{\rm dom}$,
$W_{\rm dom}$, the exact total moment, and hence $f_{\rm dom}$ and the
aggregate deficit, are meaningful. Those aggregates are unaffected,
since the dominant line is converged and the total moments are exact.

\subsection{Why $N_s=8$ is not used}

A two-cell cluster cannot serve as a second data point, and the reason
is kinematic rather than statistical. Its levels are
$\{-3.6,-2.4,-1.6,-0.4,+0.4,+1.6,+2.4,+3.6\}$, so at
$N_\uparrow=N_\downarrow=4$ every bonding--antibonding pair is either
completely filled or completely empty. The cluster is then exactly
$n_-$-dark at $U=0$: $n_-|0\rangle=0$, $\langle H_\perp\rangle=0$, and
the sum rule reads $0=0$. It possesses no noninteracting $2t_h$ anchor,
and with two $B$ sites an $A$ site cannot have three distinct $B$
neighbours in any case. It was therefore excluded from the
branch-continuity analysis. The $N_s=12$ cluster was selected because
it satisfies $z=3$ with distinct neighbours, no bond multiplicity, and
the correct $U=0$ anchor --- the properties the comparison requires.

\end{document}